\documentclass[
draft
,numberedheadings 
,varioref
]
{aipproc}

\layoutstyle{8x11single}

\usepackage{amsmath}
\usepackage{amssymb}

\usepackage[english]{babel}
\usepackage[utf8]{inputenc}

\usepackage[position=bottom,justification=centering]{subfig}
\usepackage{graphicx}

\usepackage[bbgreekl]{mathbbol}
\usepackage{bm} 
\usepackage{MnSymbol} 
\usepackage{gensymb}

\usepackage{units}
\usepackage{tensor}
\usepackage{accents}

\usepackage{booktabs}

\usepackage{todonotes}

\usepackage{xcolor}
\usepackage{cases}

\DeclareMathOperator{\divergence}{div}

\DeclareMathOperator{\Tr}{Tr}

\newcommand{\reference}{\mathrm{ref}}

\newcommand{\bydefinition}{\mathrm{def}}
\newcommand{\traceless}[1]{{#1}_{\delta}}

\newcommand{\diff}{\mathrm{d}}
\newcommand{\Diff}[1][]{\mathrm{D}_{#1}} 

\renewcommand{\vec}[1]{\ensuremath{\mathbf{#1}}}

\makeatletter
\@ifpackageloaded{bm}%
{\renewcommand{\vec}[1]{\ensuremath{\bm{#1}}}%
}{%
\relax
}
\makeatother
\newcommand{\tensorq}[1]{\ensuremath{\mathbb{#1}}}      
\newcommand{\tensorc}[1]{\ensuremath{\mathrm{#1}}}      

\newcommand{\transpose}[1]{#1^\top}

\newcommand{\inverse}[1]{#1^{-1}}

\newcommand{\identity}{\ensuremath{\tensorq{I}}}

\newcommand{\cstress}{\tensorq{T}}

\newcommand{\fgrad}{\tensorq{F}}

\newcommand{\rcg}{\tensorq{C}}

\newcommand{\lcg}{\tensorq{B}}

\makeatletter
\@ifpackageloaded{bm}%
{%
}{%

}

\@ifpackageloaded{bm}%
{%
 
}{%

}

\@ifpackageloaded{bm}%
{%
}{%

}
\makeatother

\newcommand{\generictensor}{{\tensorq{A}}}
\newcommand{\generictensorc}{\tensorc{A}} 

\newcommand{\gradasym}{\ensuremath{\tensorq{W}}}
\newcommand{\gradsym}{\ensuremath{\tensorq{D}}}

\newcommand{\gradvl}{\ensuremath{\tensorq{L}}}

\newcommand{\kdelta}[1]{\tensor{\delta}{#1}}


\newcommand{\ienergy}{\ensuremath{e}} 
\newcommand{\fenergy}{\ensuremath{\psi}} 
\newcommand{\entropy}{\ensuremath{\eta}} 

\newcommand{\temp}{\ensuremath{\theta}} 
\newcommand{\mns}{\ensuremath{m}} 

\newcommand{\nettenergy}{\ensuremath{E}_{\mathrm{tot}}} 
\newcommand{\netentropy}{\ensuremath{S}} 

\newcommand{\cheatvol}{\ensuremath{c_{\mathrm{V}}}}

\newcommand{\efluxc}{\vec{j}_{e}} 

\newcommand{\entfluxc}{\vec{j}_{\entropy}} 

\newcommand{\entprodc}{\xi} 

\newcommand{\pd}[2]{\ensuremath{\frac{\partial {#1}}{\partial {#2}}}}
\newcommand{\ppd}[2]{\ensuremath{\frac{\partial^2 {#1}}{\partial {#2^2}}}}
\newcommand{\dd}[2]{\ensuremath{\frac{\diff {#1}}{\diff {#2}}}}

\newcommand{\fid}[1]{\ensuremath{\accentset{\triangledown}{#1}}}

\newcommand{\gfid}[1]{\ensuremath{\accentset{\medsquare}{#1}}}

\makeatletter
\@ifpackageloaded{tensor}
{

}{%

}
\makeatother

\makeatletter
\@ifpackageloaded{tensor}
{

}{%

}
\makeatother

\newcommand{\absnorm}[1]{\ensuremath{\left|#1\right|}}

\makeatletter
\@ifundefined{volume}{%
}%
{%
}
\makeatother

\newcommand{\cvolumee}{\diff \mathrm{v}}

\newcommand{\csurfacees}{\diff \mathrm{s}}

\makeatletter
\@ifpackageloaded{MnSymbol} 
{
\newcommand{\tensordot}[2]{\ensuremath{#1 \vdotdot #2}} 
}{%
\newcommand{\tensordot}[2]{\ensuremath{#1 : #2}} 
}
\makeatother
\newcommand{\vectordot}[2]{\ensuremath{#1 \bullet #2}}

\newcommand{\entropyrellp}{{\mathcal S}}

\newcommand{\energyrellp}{{\mathcal E}}

\newcommand{\tempref}{\ensuremath{\temp_{\reference}}}

\newcommand{\fgradrc}{\ensuremath{\fgrad}} 

\newcommand{\gradvlI}{\ensuremath{\gradvl_1}}
\newcommand{\fgradI}{\ensuremath{\fgrad_1}}

\newcommand{\gradvlII}{\ensuremath{\gradvl_2}}
\newcommand{\fgradII}{\ensuremath{\fgrad_2}}
\newcommand{\lcgII}{\ensuremath{\lcg_{\mathrm{2}}}}
\newcommand{\rcgII}{\ensuremath{\rcg_{\mathrm{2}}}}

\newcommand{\gradvlGS}{\ensuremath{\gradvl_{\mathrm{GS}}}}
\newcommand{\fgradGS}{\ensuremath{\fgrad_{\mathrm{GS}}}}

\newcommand{\gradvlGSI}{\ensuremath{\gradvl_{1, \mathrm{GS}}}}
\newcommand{\fgradGSI}{\ensuremath{\fgrad_{1, \mathrm{GS}}}}
\newcommand{\gradsymGSI}{\ensuremath{\gradsym_{1, \mathrm{GS}}}}

\newcommand{\gradvlGSII}{\ensuremath{\gradvl_{2, \mathrm{GS}}}}
\newcommand{\fgradGSII}{\ensuremath{\fgrad_{2, \mathrm{GS}}}}
\newcommand{\lcgGSII}{\ensuremath{\lcg_{\mathrm{2}, \mathrm{GS}}}}
\newcommand{\rcgGSII}{\ensuremath{\rcg_{\mathrm{2}, \mathrm{GS}}}}

\makeatletter
\@ifpackageloaded{MnSymbol} 
{
  \newcommand{\tensorddot}[2]{\ensuremath{#1 \vdots #2}} 
}{%
  \newcommand{\tensorddot}[2]{\ensuremath{#1 \vdots #2}} 
}
\makeatother

\usepackage{placeins}

\let\cite\citet

\usepackage{textcase}
\usepackage[tablename=TABLE, figurename=FIGURE, justification=centerlast]{caption} 
\DeclareCaptionTextFormat{up}{\MakeTextUppercase{#1}}
\usepackage{indentfirst} 
\usepackage{titlecaps}   
\renewcommand\AIPsubsubsectionformat[1]   {\centering\titlecap{#1}}
\renewcommand\AIPsubsectionformat[1]   {\centering\titlecap{#1}}
\renewcommand\AIPsectionpreskip  {\bodytextbaselineskip plus 3pt minus 1pt}
\renewcommand\AIPsubsectionpreskip  {\bodytextbaselineskip plus 3pt minus 1pt}
\renewcommand\AIPsubsubsectionpreskip  {\bodytextbaselineskip plus 3pt minus 1pt}
\Addlcwords{and of the in on with via at to for a}

\usepackage{letltxmacro}
\LetLtxMacro{\originaleqref}{\ref} 
\renewcommand{\eqref}{Eq.~\originaleqref}

\usepackage{mdframed}

\usepackage{ragged2e}

\newcommand{\SP}{\texttt{\char`\ }} 
\newcommand{\showspaces}[1]{\begingroup\let\ =\SP #1\endgroup}

\begin{document}

\title{\titlecap{On diffusive variants of some classical viscoelastic rate-type models}}


\author{Mark Dostalík\textsuperscript{\small 1, a)}}{
  address={%
\textsuperscript{\small \textup{1}}Faculty of Mathematics and Physics, Charles University, Sokolovsk\'a 83, Praha 8 -- Karl\'{\i}n, CZ~186~75, \hbox{Czech Republic}\\
\textsuperscript{\small \textup{2}}Institute of Science and Technology Austria, Am Campus 1, Klosterneuburg, A~3400, Austria\\
\bigskip
{
\normalfont
\textsuperscript{a)}dostalik@karlin.mff.cuni.cz\\
\textsuperscript{b)}Corresponding author: prusv@karlin.mff.cuni.cz\\
\textsuperscript{c)}tomas.skrivan@ist.ac.at
}
}
}

\author{V\'{\i}t Pr\r{u}\v{s}a\textsuperscript{\small 1, b)}}{
}

\author{Tom\'a\v{s} Sk\v{r}ivan\textsuperscript{\small 2, c)}}{
}

\begin{abstract}
We present a thermodynamically based approach to the design of models for viscoelastic fluids with stress diffusion effect. In particular, we show how to add a stress diffusion term to some standard viscoelastic rate-type models (Giesekus, FENE-P, Johnson--Segalman, Phan-Thien--Tanner and Bautista--Manero--Puig) so that the resulting models with the added stress diffusion term are thermodynamically consistent in the sense that they obey the first and the second law of thermodynamics.  We point out the potential applications of the provided thermodynamical background in the study of flows of fluids described by the proposed models. 
\end{abstract} 

\maketitle


\begin{mdframed}[hidealllines=true, backgroundcolor=yellow]
  The manuscript has been published as Dostal\'{\i}k, M. and Pr\r{u}\v{s}a, V.  and Sk\v{r}rivan, T.: On diffusive variants of some classical viscoelastic rate-type models, AIP Conference Proceedings, 2107, 1, 020002, 2019, doi:10.1063/1.5109493. Unfortunately, the published version of this manuscript contains several misprints.
  \begin{itemize}
  \item Misprint in the specification of the additive constant in the Helmholtz free energy \emph{ansatz} for the FENE-P model. The correct treatment of the FENE-P is discussed at the end of the manuscript is section ``Helmholtz free energy for the FENE-P model''. 
  \item  Misprint in the specification of the entropy production for the Giesekus model. The correct treatment of the entropy production for the Giesekus model is discussed at the end of the manuscript is section ``Entropy production for the Giesekus model''. 
  \end{itemize}
  We are fully responsible for the misprints. 
\end{mdframed}

\tableofcontents 

\section{Introduction}
\label{sec:introduction}
The behaviour of complex fluids such as wormlike micellar solutions is on the \emph{macroscopic level} often described using viscoelastic rate-type models with an added stress diffusion term, see~\cite{cates.me.fielding.sm:rheology}, \cite{subbotin.av.malkin.ay.ea:self-organization} or \cite{fardin.m.radulescu.o.ea:stress}. The derivation of most of these phenomenological models usually proceeds in a rather \emph{ad hoc} manner. One takes a well established viscoelastic rate-type model, and adds a stress diffusion term to the evolution equation for the extra stress tensor, while the stress diffusion term takes the form of Laplace operator acting on the extra stress tensor. In this way one obtains the \emph{diffusive Johnson--Segalman model}, see for example~\cite{olmsted.pd.radulescu.o.ea:johnson-segalman}, out of the standard Johnoson--Segalman model, see~\cite{johnson.mw.segalman.d:model}. Similarly the \emph{diffusive Giesekus model}, see for example~\cite{helgeson.me.vasquez.pa.ea:rheology,helgeson.me.reichert.md.ea:relating} or \cite{cheng.p.burroughs.mc.ea:distinguishing}, is built out of the standard Giesekus model, see~\cite{giesekus.h:simple}, or the \emph{diffusive Rolie--Poly model}, see for example \cite{adams.jm.fielding.sm.ea:transient}, \cite{chung.c.uneyama.t.ea:numerical} or \cite{carter.ka.girkin.jm.ea:shear}, is built out of the standard Rolie--Poly model, see~\cite{likhtman.ae.graham.rs:simple}, and so forth. The presence of the stress diffusion term is motivated by several arguments.

\emph{First}, the stress diffusion term naturally arises in the analysis of microscopic dumbbell models. If the presumably \emph{strong spatial inhomogeneity} of the flow field is properly taken into account, see~\cite{el-kareh.aw.leal.lg:existence}, then a stress diffusion term appears in the averaged governing equations on the macroscopic level. (Here the term ``diffusion'' simply refers to the fact that the stress diffusion term is given by the Laplace operator.) \emph{Second}, the addition of the stress diffusion term to the standard viscoelastic models has pleasing implications regarding the flow dynamics, in particular in the study of the \emph{shear banding} phenomenon, see~\cite{fardin.m.radulescu.o.ea:stress}, \cite{divoux.t.fardin.ma.ea:shear} and \cite{olmsted.pd.radulescu.o.ea:johnson-segalman}. \emph{Third}, the presence of the stress diffusion term is convenient from a mathematical point of view since it provides a regularisation term in the equations, see~for example~\cite{thomases.b:analysis}, \cite{barrett.jw.boyaval.s:existence}, \cite{chupin.l.martin.s:stationary}, \cite{chupin.l.ichim.a.ea:stationary} or \cite{lukacov-medvidova.m.notsu.h.ea:energy}.

This \emph{ad hoc} approach might be acceptable provided that one is interested in the evolution of mechanical fields (pressure, velocity, extra stress) only, and if one is willing to completely ignore the temperature field. If the evolution of mechanical fields and thermal field is coupled, for example via the temperature dependence of the stress diffusion coefficient, which could be the case in practice, see~\cite{mohammadigoushki.h.muller.sj:flow}, then a different approach to the model design must be taken. Namely thermodynamically consistent models describing the evolution of mechanical fields as well as the thermal field must be developed.

While a thermodynamical background is usually available for the standard viscoelastic models, see for example~\cite{wapperom.p.hulsen.ma:thermodynamics}, \cite{dressler.m.edwards.bj.ea:macroscopic}, \cite{rajagopal.kr.srinivasa.ar:thermodynamic}, \cite{ellero.m.espanol.p.ea:thermodynamically} or~\cite{pavelka.m.klika.v.ea:multiscale} for various approaches, much less is known about thermodynamically consistent models describing complex fluids such as wormlike micellar solutions. The existing thermodynamically consistent models are based on the modelling of internal structure of the fluid either by the means of a single internal parameter, see for example~\cite{manero.o.perez-lopez.jh.ea:thermodynamic}, or by the means of multiple-species reactions in the spirit of the model introduced by~\cite{vasquez.pa.mckinley.gh.ea:network}, see for example thermodynamical analysis by~\cite{grmela.m.chinesta.f.ea:mesoscopic} or \cite{germann.n.cook.lp.ea:nonequilibrium}.

On the other hand, \emph{thermodynamical analysis of models that contain the stress diffusion term} in the form of Laplace operator has not been provided, to our best knowledge, until very recently, see~\cite{malek.j.prusa.v.ea:thermodynamics}. In their work \cite{malek.j.prusa.v.ea:thermodynamics} have followed the purely phenomenological thermodynamical approach to viscoelastic rate-type models introduced by \cite{rajagopal.kr.srinivasa.ar:thermodynamic}, see also~\cite{malek.j.rajagopal.kr.ea:on}, \cite{hron.j.milos.v.ea:on} or~\cite{malek.j.prusa.v:derivation}, and they have derived Maxwell/Oldroyd-B type models with the stress diffusion term both for compressible and incompressible fluids. \emph{In what follows we extend their analysis, and we derive diffusive variants of the Giesekus, FENE-P, Johnson--Segalman, Phan-Thien--Tanner and Bautista--Manero--Puig models}.

The basic premise of the phenomenological thermodynamical approach by \cite{rajagopal.kr.srinivasa.ar:thermodynamic}, see also~\cite{rajagopal.kr.srinivasa.ar:on*7}, is that material of interest is characterised by the way it \emph{stores the energy} and by the way it \emph{produces the entropy}. Once these mechanisms are identified then the evolution equations simply follow from this identification and the underlying kinematical assumptions.

In their analysis \cite{malek.j.prusa.v.ea:thermodynamics} have explored the possibility to interpret the stress diffusion effect in both ways. Namely, the stress diffusion term has been identified either as a symptom of a \emph{nonstandard energy storage mechanism} or a \emph{nonstandard entropy production mechanism}. In both cases one obtains a stress diffusion term in the evolution equation for the extra stress tensor, but the final models differ in the structure of the full Cauchy stress tensor and the specific heat capacity, see~\cite{malek.j.prusa.v.ea:thermodynamics} for details.

In the present analysis we focus on the second option, that is we interpret the stress diffusion as a consequence of a nonstandard entropy producing mechanism. This is consistent with the microscopic analysis by~\cite{el-kareh.aw.leal.lg:existence}, who have shown that the stress diffusion coefficient depends on the \emph{hydrodynamic resistance} of one dumbbell bead. Consequently it is reasonable to interpret the stress diffusion as a dissipative (entropy producing) process. Concerning the characterisation of the entropy production, we in principle use the same \emph{ansatz} as that used by~\cite{malek.j.prusa.v.ea:thermodynamics}. In particular the corresponding term in the entropy production \emph{ansatz} is quadratic in the gradient of the ``extra stress'' field.

Concerning the characterisation of the energy storage mechanism, which is in the given framework done via the specification of the Helmholtz free energy, we use formulae that are (Giesekus, Johnson--Segalman, Phan-Thien--Tanner models) \emph{structurally} similar to that used by~\cite{malek.j.prusa.v.ea:thermodynamics}, or that are known for the classical models without stress diffusion, see \cite{wapperom.p.hulsen.ma:thermodynamics}, \cite{hu.d.lelievre.t:new} and also~\cite{barrett.j.boyaval.s:finite}. However, contrary to the approach by~\cite{rajagopal.kr.srinivasa.ar:thermodynamic} and~\cite{malek.j.prusa.v.ea:thermodynamics}, we need to make different assumptions concerning the underlying kinematics. In particular, the kinematical assumptions must be designed in such a way that they allow one to \emph{replace the standard upper convected derivative by the Gordon--Schowalter derivative}, see~\cite{gordon.rj.schowalter.wr:anisotropic}. This requires substantial changes in the analysis provided by~\cite{malek.j.prusa.v.ea:thermodynamics}.

Once the specification of the Helmholtz free energy (energy storage mechanism) and the entropy production (entropy production mechanism) is done, see Table~\ref{tab:models}, and once the underlying kinematical assumptions are known, then the governing equations for the mechanical variables as well as for the temperature follow immediately. This shows the flexibility of the adopted thermodynamical approach --- the governing equations are indeed obtained by combining three simple pieces of information.   

The paper is organised as follows. First, we briefly describe the thermodynamical framework introduced by~\cite{rajagopal.kr.srinivasa.ar:thermodynamic}, and we in detail describe the underlying kinematical assumptions that allow one to incorporate the Gordon--Schowalter derivative into the models. Once the preliminary work is done, we introduce the specific Helmholtz free energy and the entropy production that lead one after the other to the diffusive variants of the Giesekus, FENE-P, Johnson--Segalman, Phan-Thien--Tanner and Bautista--Manero--Puig models, and we explicitly write down the corresponding evolution equations for the temperature field.  (The Helmholtz free energy/entropy production pairs for all the models are summarised in Table~\ref{tab:models}.) We also focus on the implied structure of the corresponding \emph{energy and entropy fluxes}. This is essential piece of information provided that one wants to precisely characterize what is in the context of complex fluids meant by thermodynamically open/isolated systems. The paper is concluded by a brief discussion of the potential use of the provided thermodynamical background in the study of stability of flows of fluids described by the proposed models. 

\begin{table}[h]
  \centering
  \begin{tabular}{lll}  
    \toprule
    Helmholtz free energy $\fenergy_m$ (mechanical part) & Entropy production $\entprodc_m$ (contribution of mechanical quantities)\\
    \midrule
    \multicolumn{2}{l}{\textbf{Giesekus}} \\[0.5em]
    $
    \frac{\mu}{2\rho} \left[ \Tr \lcgII - 3 - \ln \det \lcgII \right]
    $
                          &
                            $
                            \frac{1}{\temp}
                            \left\{
                            2 \nu \tensordot{\traceless{\gradsym}}{\traceless{\gradsym}}
                            +
                            \frac{\mu^2}{2 \nu_1} \Tr 
                            \left[ 
                            \lambda \lcgII^2 + (1 - 3 \lambda) \lcgII + (1 + \lambda) \inverse{\lcgII} + (3 \lambda - 2) \identity
                            \right]
                            +
                            \frac{\mu \tilde{\mu}}{2\nu_1}
                            \tensorddot{
                            \nabla \lcgII
                            }
                            {
                            \nabla \lcgII
                            }
                            \right\}
                            $
    \\[1em]
    \midrule
    \multicolumn{2}{l}{\textbf{FENE-P}} \\[0.5em]
    $
    \frac{\mu}{2\rho}
    \left[
    -
    b \ln \left( 1 - \frac{1}{b} \Tr \lcgII \right) 
    -
    3
    -
    \ln \det \lcgII
    \right]
    $
                          &
                            $
                            \frac{1}{\temp}
                            \left\{
                            2 \nu \tensordot{\traceless{\gradsym}}{\traceless{\gradsym}}
                            +
                            \frac{\mu^2}{2 \nu_1} 
                            \left[ 
                            \left( 1 - \frac{1}{b} \Tr \lcgII \right)^{-2} \Tr \lcgII
                            -
                            6 \inverse{\left( 1 - \frac{1}{b} \Tr \lcgII \right)}
                            +
                            \Tr \inverse{\lcgII}
                            \right]
                            +
                            \frac{\mu \tilde{\mu}}{2\nu_1}
                            \tensorddot{
                            \nabla 
                            \left[ 
                            \inverse{\left( 1 - \frac{1}{b} \Tr \lcgII \right)} \lcgII
                            \right]
                            }
                            {
                            \nabla 
                            \left[ 
                            \inverse{\left( 1 - \frac{1}{b} \Tr \lcgII \right)} \lcgII
                            \right]
                            }
                            \right\}
                            $
    \\[1em]
    \midrule
    \multicolumn{2}{l}{\textbf{Johnson--Segalman}} \\[0.5em]
    $
    \frac{\mu}{2\rho}
    \left[
    \Tr \lcgGSII
    -
    3
    -
    \ln \det \lcgGSII
    \right]
    $
                          &
                            $
                            \frac{1}{\temp}
                            \left\{
                            2 \nu \tensordot{\traceless{\gradsym}}{\traceless{\gradsym}}
                            +
                            \frac{\mu^2}{2 \nu_1} 
                            \left[ 
                            \Tr \lcgGSII + \Tr \inverse{\lcgGSII} - 6
                            \right]
                            +
                            \frac{\mu \tilde{\mu}}{2\nu_1}
                            \tensorddot{
                            \nabla \lcgGSII
                            }
                            {
                            \nabla \lcgGSII
                            }
                            \right\}
                             $
    \\[1em]
    \midrule
    \multicolumn{2}{l}{\textbf{Phan--Thien--Tanner}} \\[0.5em]
    $
    \frac{\mu}{2\rho}
    \left[
    \Tr \lcgGSII
    -
    3
    -
    \ln \det \lcgGSII
    \right]
    $
                          &
                            $
                            \frac{1}{\temp}
                            \left\{
                            2 \nu \tensordot{\traceless{\gradsym}}{\traceless{\gradsym}}
                            +
                            \frac{\mu^2}{2 \nu_1} f(\lcgGSII)
                            \left[ 
                            \Tr \lcgGSII + \Tr \inverse{\lcgGSII} - 6
                            \right]
                            +
                            \frac{\mu \tilde{\mu}}{2\nu_1}
                            \tensorddot{
                            \nabla \lcgGSII
                            }
                            {
                            \nabla \lcgGSII
                            }
                            \right\}
                            $
    \\[1em]
    \midrule
    \multicolumn{2}{l}{\textbf{Bautista--Manero--Puig}} \\[0.5em]
    $
    \frac{\mu}{2\rho}
    \left[
    \Tr \lcgII
    -
    3
    -
    \ln \det \lcgII
    \right]
    +
    \frac{\chi}{2 \rho}
    \left(
    \varphi_0 - \varphi
    \right)^2
    $
                          &
                            $
                            \frac{1}{\temp}
                            \left\{
                            2 \nu \tensordot{\traceless{\gradsym}}{\traceless{\gradsym}}
                            +
                            \frac{\mu^2 \varphi}{2} 
                            \left[ 
                            \Tr \lcgII + \Tr \inverse{\lcgII} - 6
                            \right]
                            +
                            \frac{\beta \mu \tilde{\mu}}{2}
                            \tensorddot{
                            \nabla \lcgII
                            }
                            {
                            \nabla \lcgII
                            }
                            +
                            \chi
                            \frac{\left( \varphi_0 - \varphi \right)^2}{\tau}
                            \right\}
                            $
    \\[1em]
    \bottomrule
  \end{tabular}
  \caption{Mechanical part of Helmholtz free energy $\fenergy_m$ and contribution of mechanical quantities to entropy production $\entprodc_m$ for some viscoelastic rate-type models with a stress diffusion term. If the stress diffusion coefficient $\tilde{\mu}$ is equal to zero, then the given Helmholtz free energy/entropy production pair leads to the corresponding classical model. The Helmholtz free energy $\fenergy$ and entropy production $\entprodc$ are obtained from the tabulated formulae via
    $
    \fenergy
    =_{\bydefinition}
    -
    \cheatvol^{\mathrm{NSE}}
    \temp
    \left(
      \ln \frac{\temp}{\tempref} - 1
    \right)
    +
    \fenergy_m
    $
    and
    $
    \entprodc = \entprodc_m + \kappa \frac{\absnorm{\nabla \temp}^2}{\temp^2}
    $.
    \bigskip}
  \label{tab:models}
\end{table}

\FloatBarrier

\section{Preliminaries}
\label{sec:preliminaries}

The methodology introduced by~\cite{rajagopal.kr.srinivasa.ar:thermodynamic} is purely phenomenological one in the sense that it does not rely on the knowledge of the internal microscopic structure of the given fluid. In particular, it does not operate with the notion of the conformation tensor. The additional tensorial variable arises as a consequence of the virtual decomposition of the total deformation to elastic and dissipative response, see Figure~\ref{fig:viscoelastic-kinematics-a}. This decomposition clearly mimics the standard spring--dashpot analogue for Maxwell type viscoelastic fluid, see for example~\cite{wineman.as.rajagopal.kr:mechanical}.

If necessary, this decomposition can easily be generalised in order to design models that correspond to more involved spring-dashpot analogues such as Burgers model, see~\cite{karra.s.rajagopal.kr:development,karra.s.rajagopal.kr:thermodynamic} or~\cite{malek.j.rajagopal.kr.ea:derivation}, which is a model that is popular in the modelling of geomaterials and biological fluids, see for example~\cite{hron.j.rajagopal.kr.ea:flow}, \cite{malek.j.rajagopal.kr.ea:thermodynamically,malek.j.rajagopal.kr.ea:thermodynamically*1} or~\cite{tuma.k.stein.j.ea:motion}. However, in order to get viscoelastic rate-type models with the Gordon--Schowalter derivative instead of the upper convected derivative, one needs to rethink the decomposition. 

\subsection{Kinematics}
\label{sec:kinematics}

\begin{ltxfigure}[h]
  \centering
  \subfloat[\label{fig:viscoelastic-kinematics-a}Decomposition of deformation gradient to a dissipative and elastic part.]{\includegraphics[width=0.4\textwidth]{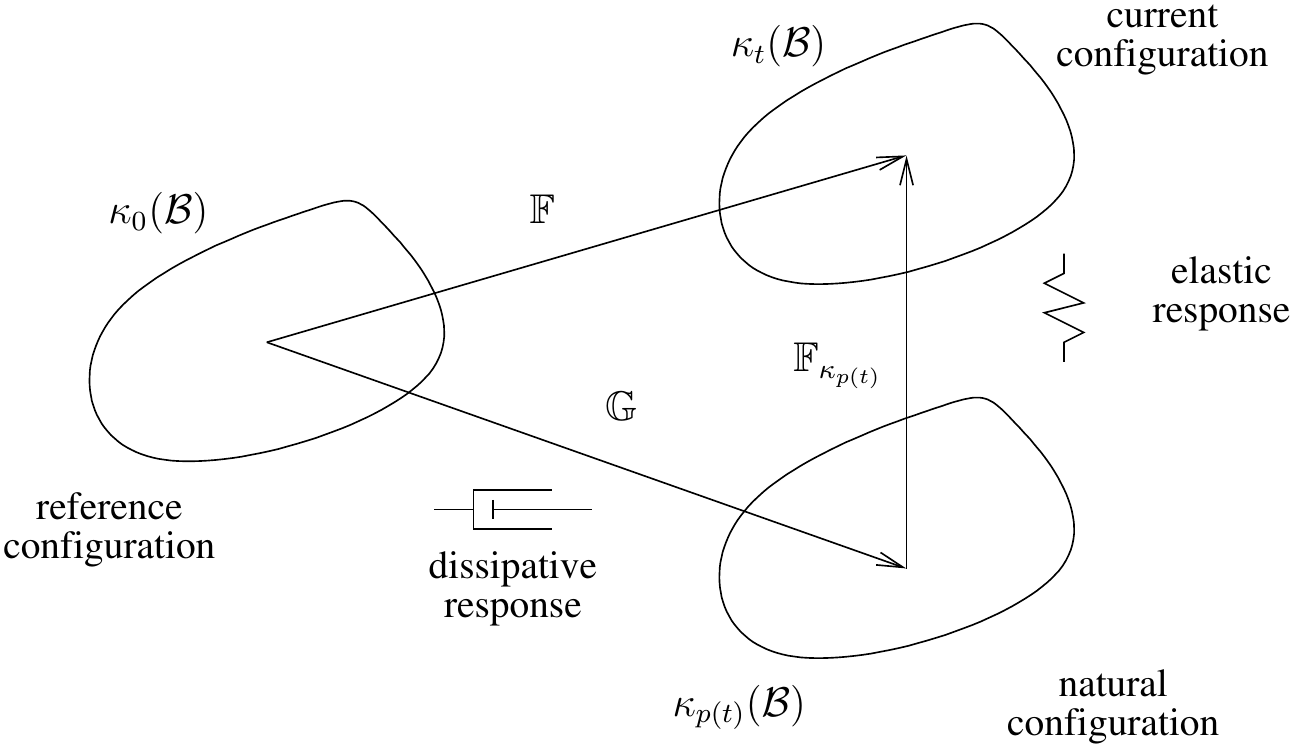}}
  \qquad
  \subfloat[\label{fig:viscoelastic-kinematics-b}General decomposition of deformation gradient.]{\includegraphics[width=0.4\textwidth]{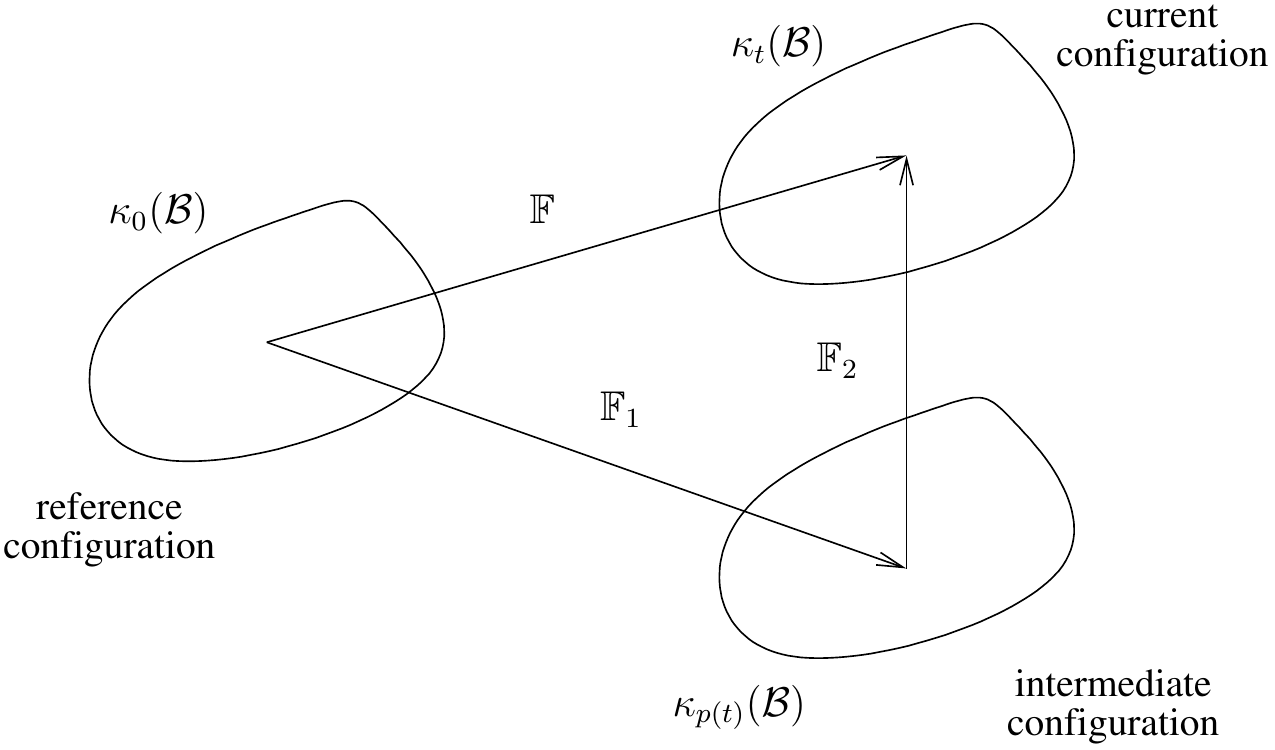}}
  \caption{Viscoelastic fluid -- kinematics.}
  \label{fig:viscoelastic-kinematics}
\end{ltxfigure}

\subsubsection{Models with upper convected derivative}
\label{sec:models-with-upper}
Let us now consider a general decomposition as shown in Figure~\ref{fig:viscoelastic-kinematics-b}. If the total deformation is seen as a composition of the two deformations, then the total deformation gradient $\fgrad$ can be written~as
\begin{equation}
  \label{eq:deformation-composition}
  \fgradrc = \fgradII \fgradI,
\end{equation}
where $\fgradII$ and $\fgradI$ are the deformation gradients of the partial deformations. (Note that such a decomposition is frequent in other settings such as plasticity, see for example~\cite{gurtin.me.fried.e.ea:mechanics}.) Motivated by the standard relation
\begin{equation}
  \label{eq:flf-identity}
  \dd{\fgrad}{t} = \gradvl \fgrad
\end{equation}
between the spatial velocity gradient $\gradvl=_{\bydefinition} \nabla \vec{v}$ and the deformation gradient $\fgrad$, where
\begin{equation}
  \label{eq:material-time-derivative}
  \dd{}{t}=_{\bydefinition} \pd{}{t} + \vectordot{\vec{v}}{\nabla},
\end{equation}
denotes the material time derivative, we introduce new tensorial rate quantities $\gradvlI$ and $\gradvlII$, and also symmetric and skew-symmetric parts of these tensors. Their definitions along with the corresponding symmetric and asymmetric parts read
\begin{subequations}
  \label{eq:gradvl-gradsym-gradasym}
  \begin{align}
    \gradvl &= \dd{\fgrad}{t} \inverse{\fgrad},
    &
    \gradvlI &=_{\bydefinition} \dd{\fgradI}{t} \inverse{\fgradI},
    &
    \gradvlII &=_{\bydefinition} \dd{\fgradII}{t} \inverse{\fgradII}
    \\
    \gradsym &= _{\bydefinition} \frac{1}{2} \left(\gradvl + \transpose{\gradvl}\right),
    &
    \gradsym_1 &= _{\bydefinition} \frac{1}{2} \left(\gradvlI + \transpose{\gradvlI}\right),
    &
    \gradsym_2 &= _{\bydefinition} \frac{1}{2} \left(\gradvlII + \transpose{\gradvlII}\right),
    \\
    \gradasym &= _{\bydefinition} \frac{1}{2} \left(\gradvl - \transpose{\gradvl}\right),
    &
    \gradasym_1 &= _{\bydefinition} \frac{1}{2} \left(\gradvlI - \transpose{\gradvlI}\right),
    &
    \gradasym_2 &= _{\bydefinition} \frac{1}{2} \left(\gradvlII - \transpose{\gradvlII}\right).
    \end{align}
\end{subequations}
Moreover, we define the right and left Cauchy--Green tensors associated to the partial deformation associated with the deformation gradient $\fgradII$,
\begin{equation}
  \label{eq:lcg-rcg}
  \rcgII =_{\bydefinition} \transpose{\fgradII} \fgradII,
  \qquad
  \lcgII =_{\bydefinition} \fgradII \transpose{\fgradII}.
\end{equation}
By taking the material time derivative of \eqref{eq:deformation-composition}, we obtain
\begin{equation}
  \label{eq:gradvl-identity}
  \gradvl
  =
  \gradvlII + \fgradII \gradvlI \inverse{\fgradII},
\end{equation}
which then yields a kinematical identity
\begin{equation}
  \label{eq:fid-lcg_2}
  \fid{\overline{\lcgII}} 
  =   
  -
  2 \fgradII \gradsym_1 \transpose{\fgradII},
\end{equation}
where
\begin{equation}
  \label{eq:upper-convected-derivative}
  \fid{\generictensor}
  =_{\bydefinition} 
  \dd{\generictensor}{t} 
  - 
  \gradvl \generictensor 
  - 
  \generictensor \transpose{\gradvl}
  .
\end{equation}
denotes the upper convected derivative. In the standard derivation by~\cite{rajagopal.kr.srinivasa.ar:thermodynamic} one then uses thermodynamical arguments and writes the right hand side of~\eqref{eq:fid-lcg_2} as a function of $\lcgII$.

\subsubsection{Models with Gordon{--}Schowalter derivative}
\label{sec:models-with-gordon}
The evolution equations for both the Johnson--Segalman model, see~\cite{johnson.mw.segalman.d:model}, and the Phan--Thien--Tanner model, see~\cite{phan-thien.n.tanner.ri:new}, however contain the so-called Gordon--Schowalter derivative
\begin{equation}
  \label{eq:gordon-schowalter-derivative}
  \gfid{\generictensor}
  =_{\bydefinition}
  \dd{\generictensor}{t} 
  - 
  a 
  \left( 
    \gradsym \generictensor + \generictensor \gradsym 
  \right)
  -
  \left( 
    \gradasym \generictensor + \generictensor \transpose{\gradasym}
  \right)
  ,
\end{equation}
where $a \in [-1, 1]$, see~\cite{gordon.rj.schowalter.wr:anisotropic}. (Note that if $a=1$, then the Gordon--Schowalter derivative $\gfid{\generictensor}$ reduces to the standard upper-convected derivative $\fid{\generictensor}$.) This type of derivative presumably takes into account ``non-affine'' molecular motion, see~\cite{johnson.mw.segalman.d:model}. Consequently, we would like to derive a kinematical formula similar to~\eqref{eq:fid-lcg_2} for the Gordon--Schowalter derivative. The formula \eqref{eq:gordon-schowalter-derivative} can be rewritten as
\begin{equation}
  \label{eq:gordon-schowalter-derivative-gen}
  \gfid{\generictensor}
  =
  \dd{\generictensor}{t} 
  - 
  \gradvlGS \generictensor 
  - 
  \generictensor \transpose{\gradvlGS}
  ,
\end{equation}
where we have defined
\begin{equation}
  \label{eq:gradvlGS}
  \gradvlGS
  =_{\bydefinition}
  a \gradsym + \gradasym
  =
  \gradvl + (a - 1) \gradsym.
\end{equation}
Note that when $a=1$ the tensor $\gradvlGS$ reduces to the standard velocity gradient $\gradvl$. Motivated by \eqref{eq:flf-identity} we then define tensor~$\fgradGS$ as the solution of the differential equation
\begin{equation}
  \label{eq:fgradGS-definition}
  \dd{\fgradGS}{t} = \gradvlGS \fgradGS,
  \qquad
  \left. \fgradGS \right|_{t=0} = \identity.
\end{equation}
Similarly to \eqref{eq:gradvlGS} we define
\begin{equation}
  \label{eq:gradvlGSII-definition}
  \gradvlGSII
  =_{\bydefinition}
  a \gradsym_2 + \gradasym_2
  =
  \gradvlII + (a - 1) \gradsym_2,
\end{equation}
and then define $\fgradGSII$ as the solution of the differential equation
\begin{equation}
  \label{eq:fgradGSII-definition}
  \dd{\fgrad_{2, \mathrm{GS}}}{t} = \gradvlGSII \fgrad_{2, \mathrm{GS}},
  \qquad
  \left. \fgrad_{2, \mathrm{GS}} \right|_{t=0} = \identity.
\end{equation}
As in the previous case we define the right and left Cauchy--Green tensors corresponding to the tensorial quantity $\fgrad_{2, \mathrm{GS}}$
\begin{equation}
  \label{eq:lcgGSII-rcgGSII}
  \rcgGSII =_{\bydefinition} \transpose{\fgradGSII} \fgradGSII,
  \qquad
  \lcgGSII =_{\bydefinition} \fgradGSII \transpose{\fgradGSII}.
\end{equation}
We complete the definitions of the new tensorial quantities by defining $\gradvlGSI$ and $\fgradGSI$ as%
\footnote{
  The invertibility of the tensor $\fgradGS$ follows from the fact that
  \begin{equation}
  \label{eq:det-fgradgen-ode}
  \dd{}{t} \det \fgradGS 
  = 
  \left( 
    \det \fgradGS
  \right)
  \Tr 
  \left( 
    \dd{\fgradGS}{t} \inverse{\fgradGS} 
  \right)
  =
  \left( 
    \det \fgradGS
  \right)
  \Tr \gradvlGS
  =
  a
  \left( 
    \det \fgradGS
  \right)
  \Tr \gradvl
  .
  \end{equation}
  We see that the solution of the differential equation \eqref{eq:det-fgradgen-ode}, subject to the initial condition $\left. \det \fgradGS \right|_{t=0} = 1$, reads $\det \fgradGS (t) = \exp \left( \int_0^t a \Tr \gradvl(\tau) \, \mathrm{d} \tau \right)$. Hence, $\fgradGS$ remains invertible at all times. Similarly, we could show that the tensor $\fgradGSII$ remains invertible at all times and thus \eqref{eq:fgradGSI} then yields that $\fgradGSI$ is invertible as well.
}
\begin{subequations}
  \begin{align}
    \label{eq:fgradGSI}
    \fgradGSI
    &=_{\bydefinition}
    \inverse{\fgradGSII} \fgradGS,
    \\
    \label{eq:gradvlGSI}
    \gradvlGSI
    &=_{\bydefinition}
    \dd{\fgradGSI}{t} \inverse{\fgradGSI}.
  \end{align}
\end{subequations}

Definition \eqref{eq:fgradGSI} yields
\begin{equation}
  \label{eq:deformationGS-composition}
  \fgradGS = \fgrad_{2, \mathrm{GS}} \fgradGSI,
\end{equation}
which is an analogue of the formula \eqref{eq:deformation-composition}. By taking the material time of \eqref{eq:deformationGS-composition}, we obtain
\begin{equation}
  \label{eq:gradvlGS-identity}
  \gradvlGS
  =
  \gradvlGSII + \fgradGSII \gradvlGSI \inverse{\fgradGSII},
\end{equation}
which finally yields the kinematical identity
\begin{equation}
  \label{eq:gfid-lcgGSII}
  \gfid{\overline{\lcgGSII}} 
  =   
  -
  2 \fgrad_{2, \mathrm{GS}} \gradsymGSI \transpose{\fgrad_{2, \mathrm{GS}}},
\end{equation}
where
\begin{equation}
  \label{eq:gradsymGSI-definition}
  \gradsymGSI
  =_{\bydefinition}
  \frac{1}{2}
  \left(
    \gradvlGSI + \transpose{\gradvlGSI}
  \right)
  .
\end{equation}
Formula \eqref{eq:gfid-lcgGSII} is an analogue of the kinematical identity~\eqref{eq:fid-lcg_2}. Further, one can easily show that
\begin{subequations}
  \label{eq:time-derivatives-knematical-identities}
  \begin{align}
    \label{eq:1}
    \dd{}{t} \Tr \lcgGSII 
    &= 
    2 a \tensordot{\lcg_{2, \mathrm{GS}}}{\gradsym} - 2 \tensordot{\rcgGSII}{\gradsymGSI},
    \\
    \label{eq:2}
    \dd{}{t} \ln \det \lcgGSII 
    &=
      2 \tensordot{\identity}{\gradsym}
      -
      2 \tensordot{\identity}{\gradsymGSI},
  \end{align}
\end{subequations}
where $\tensordot{\generictensor}{\tensorq{B}} = _{\bydefinition} \Tr \left( \generictensor \transpose{\tensorq{B}} \right)$ denotes the standard scalar product on the space of matrices. (Note that the first term on the right-hand-side of~\eqref{eq:2} vanishes for the isochoric deformation $\divergence \vec{v} = 0$.) Moreover, using the kinematical identity~\eqref{eq:gfid-lcgGSII}, we can notice that
\begin{subequations}
  \label{eq:7}
  \begin{align}
    \label{eq:9}
    \tensordot{\rcgGSII}{\gradsymGSI} &= -\frac{1}{2}\tensordot{\gfid{\overline{\lcgGSII}}}{\identity}, \\
    \label{eq:8}
    \tensordot{\identity}{\gradsymGSI} &= -\frac{1}{2}\tensordot{\gfid{\overline{\lcgGSII}}}{\inverse{\lcgGSII}},
  \end{align}
\end{subequations}
which allows us to rewrite the right hand side of~\eqref{eq:time-derivatives-knematical-identities} completely in terms of $\lcgGSII$ and $\gradsym$.

\subsection{Evolution equation for entropy}
\label{sec:general-entropy}
Once we specify the specific Helmholtz free energy $\fenergy$ (Helmholtz free energy per unit mass, $[\fenergy] = \unitfrac{J}{kg}$), we can use the general evolution equation for the specific internal energy~$\ienergy$ of a continuous medium,
\begin{equation}
  \label{eq:10}
  \rho \dd{\ienergy}{t} = \tensordot{\cstress}{\gradsym} - \divergence \efluxc,
\end{equation}
see for example~\cite{gurtin.me.fried.e.ea:mechanics}, and we can derive the evolution equation for the specific entropy $\entropy$. (Here $\rho$ denotes the density, $\cstress$ denotes the Cauchy stress tensor and $\efluxc$ represents the non-mechanical contribution to the energy flux.) Indeed, using the chain rule, the standard set of thermodynamical identities such as~$\fenergy = \ienergy - \temp \entropy$, and the assumption that $\fenergy = \fenergy(\temp, \Tr \lcgGSII, \ln \det \lcgGSII, \varphi)$, one arrives at an expression for the time derivative of the internal energy in terms of the time derivative of the entropy $\entropy$ and the time derivative of $\lcgGSII$ and $\varphi$,
\begin{equation}
  \label{eq:17}
  \dd{\ienergy}{t}
  =
  \temp
  \dd{\entropy}{t}
  +
  \pd{\fenergy}{\Tr \lcgGSII}
  \dd{}{t}\left( \Tr \lcgGSII \right)
  +
  \pd{\fenergy}{\ln \det \lcgGSII}
  \dd{}{t}
  \left(
    \ln \det \lcgGSII
  \right)
  +
  \pd{\fenergy}{\varphi}\dd{\varphi}{t}
  ,
\end{equation}
see~\cite{hron.j.milos.v.ea:on} or~\cite{malek.j.prusa.v:derivation} for details concerning a similar manipulation. In~\eqref{eq:17} the symbol~$\temp$ denotes the temperature and $\varphi$ denotes a scalar quantity. (We will make use of $\varphi$ in the discussion of thermodynamics of Bautista--Manero--Puig type model.) Using~\eqref{eq:17} on the left-hand-side of~\eqref{eq:10} then yields the sought evolution equation for the entropy~$\entropy$.

In particular, for all the models studied in the present contribution we consider the following \emph{ansatz} for the specific Helmholtz free energy
\begin{equation}
  \label{eq:fenergy-general}
  \fenergy 
  = 
  \fenergy_0 (\temp) 
  + 
  \fenergy_1 
  (\Tr \lcgGSII, \ln \det \lcgGSII)
  +
  \fenergy_2 (\varphi).
\end{equation}
For simplicity, the \emph{ansatz} for the Helmholtz free energy splits the free energy into the thermal part $\fenergy_0$ and the mechanical part $\fenergy_m = \fenergy_1 + \fenergy_2$. (See \cite{hron.j.milos.v.ea:on} for more complicated formulae for the Helmholtz free energy.) The thermal part takes---in the simplest case of a fluid with a constant specific heat capacity---the form
\begin{equation}
  \label{eq:11}
  \fenergy_0
  =
  -
  \cheatvol^{\mathrm{NSE}}
  \temp
  \left(
    \ln \frac{\temp}{\tempref} - 1
  \right)
  ,
\end{equation}
where $\tempref$ is a reference temperature and $\cheatvol^{\mathrm{NSE}}$ is a positive material parameter (specific heat at constant volume). The postulated form of the Helmholtz free energy then yields an evolution equation for the specific entropy $\entropy$ in the form
\begin{multline}
  \label{eq:entropy-evolution-from-fenergy}
  \rho \dd{\entropy}{t} 
  + 
  \divergence \left( \frac{\efluxc}{\temp} \right)
  =
  \frac{1}{\temp}
  \left\{
    \tensordot{
      \left[
      \traceless{\cstress}
      -
      2 \rho \pd{\fenergy}{\Tr \lcgGSII} a \traceless{\left( \lcgGSII \right)}
      \right]
    }{
      \traceless{\gradsym}
    }
  \right.
  \\
  \left.
    - 
    \frac{1}{2}
    \Tr
    \left[
      \gfid{\overline{\lcgGSII}}
      \left(
        2 \rho \pd{\fenergy}{\Tr \lcgGSII} \identity
        +
        2 \rho \pd{\fenergy}{\ln \det \lcgGSII} \inverse{\lcgGSII}
      \right)
    \right]
    -
    \rho \pd{\fenergy}{\varphi} \dd{\varphi}{t}
    -
    \frac{\vectordot{\efluxc}{\nabla \temp}}{\temp}
  \right\},
\end{multline}
where we have used kinematical identities~\eqref{eq:time-derivatives-knematical-identities} and~\eqref{eq:7}, and where $\traceless{\generictensor} =_{\bydefinition} \generictensor - \frac{1}{3} \left(\Tr \generictensor\right) \identity$ denotes the traceless part of the corresponding tensor.

The derived entropy evolution equation \eqref{eq:entropy-evolution-from-fenergy} that is implied by the chosen \emph{ansatz} for the Helmholtz free energy can then be ``compared'' with the general evolution equation for the entropy in a continuous medium that reads
\begin{equation}
  \label{eq:entropy-evolution-general}
  \rho \dd{\entropy}{t}
  +
  \divergence \entfluxc
  =
  \entprodc,
\end{equation}
where $\entprodc$ denotes the required (known) entropy production and $\entfluxc$ denotes the entropy flux, see~\cite{malek.j.prusa.v:derivation} for comments. This way one can identify the sought constitutive relations for $\cstress$ and $\efluxc$ and the evolution equations for~$\lcgGSII$ and~$\varphi$. Particular applications of this procedure in the derivation of specific viscoelastic rate-type models will be discussed in the following sections.

Concerning the stress diffusion term, the comparison of \eqref{eq:entropy-evolution-from-fenergy} and \eqref{eq:entropy-evolution-general} will rely on using the identity
\begin{equation}
  \label{eq:tensorddot-identity}
  c \tensorddot{\nabla \generictensor}{\nabla \generictensor}
  =
  \frac{1}{2} \divergence
  \left\{
    c \Tr
    \left[
      \left( \nabla \generictensor \right)
      \left( \generictensor - \identity \right)
      +
      \left( \generictensor - \identity \right)
      \left( \nabla \generictensor \right)
    \right]
  \right\}
  -
  \frac{1}{2} \Tr
  \left\{
    \left[
      \divergence \left( c \nabla \generictensor \right) \generictensor
      +
      \generictensor \divergence \left( c \nabla \generictensor \right)
    \right]
    \left(
      \identity - \inverse{\generictensor}
    \right)
  \right\}
  ,
\end{equation}
where $c$ is a scalar quantity and $\generictensor$ is a symmetric tensor, while the coordinate expression of the identity is 
\begin{multline}
  c
  \left[
    \pd{\tensor{\generictensorc}{_i_j}}{x_m}
    \pd{\tensor{\generictensorc}{_i_j}}{x_m}
  \right]
  =
  \frac{1}{2}
  \pd{}{x_m}
  \left[
    c
    \left(
      \pd{\tensor{\generictensorc}{_i_j}}{x_m}
      \left(
        \tensor{\generictensorc}{_j_i}
        -
        \kdelta{_j_i}
      \right)
      +
      \pd{\tensor{\generictensorc}{_l_j}}{x_m}
      \left(
        \tensor{\generictensorc}{_j_l}
        -
        \kdelta{_j_l}
      \right)
    \right)
  \right]
  \\
  -
  \frac{1}{2}
  \left[
    \pd{}{x_m}
    \left(
      c
      \pd{\tensor{\generictensorc}{_i_j}}{x_m}
    \right)
    \tensor{\generictensorc}{_j_l}
    +
    \pd{}{x_m}
    \left(
      c
      \pd{\tensor{\generictensorc}{_l_j}}{x_m}
    \right)
    \tensor{\generictensorc}{_j_i}
  \right]
  \left(
    \kdelta{_l_i}
    -
    \tensor{\left(\inverse{\generictensorc}\right)}{_l_i}
  \right)
  ,
\end{multline}
where we have used the Einstein summation convention. Note that the triple product $\tensorddot{\nabla \generictensor}{\nabla \generictensor}$ on the left-hand-side is always a nonnegative quantity.

\subsection{Evolution equation for temperature}
\label{sec:general-temperature}
Using the fact that $\entropy = - \pd{\fenergy}{\temp}$, we can rewrite the evolution equation for the entropy~\eqref{eq:entropy-evolution-general} as
\begin{equation}
  \label{eq:3}
  \rho \dd{}{t} \left( - \pd{\fenergy}{\temp} \right)
  +
  \divergence \entfluxc
  =
  \entprodc.
\end{equation}
Moreover appealing to the assumed simple structure of the Helmholtz free energy, namely to the splitting to the thermal and mechanical part, and the particular formula for the thermal part~\eqref{eq:11}, we see that
\begin{equation}
  \label{eq:4}
  \cheatvol^{\mathrm{NSE}} 
  =
  - \temp \ppd{\fenergy}{\temp}.
\end{equation}
(Note that in the case of more complicated \emph{ansatz} for the Helmholtz free energy one can get non-constant specific heat both with respect to the temperature and $\lcgGSII$, see for example~\cite{hron.j.milos.v.ea:on} or~\cite{malek.j.prusa.v.ea:thermodynamics}.) Consequently, using the chain rule we see that~\eqref{eq:3} can be rewritten as an evolution equation for the temperature in the form
\begin{equation}
  \label{eq:temperature-evolution-general}
  \rho \cheatvol^{\mathrm{NSE}} \dd{\temp}{t}
  +
  \temp \divergence
    \entfluxc
  =
  \temp \entprodc.
\end{equation}
This manipulation shows that once the Helmholtz free energy $\fenergy$ and entropy production $\entprodc$ are specified via a suitable \emph{ansatz}, then the corresponding evolution equation for temperature is a simple consequence of the choice of $\fenergy$ and $\entprodc$.

\section{Giesekus model}
\label{sec:giesekus-model}
Let us now focus on specific viscoelastic rate-type models with a stress diffusion term. First, we identify the energy storage mechanisms and the entropy production mechanisms which yield the constitutive relations and in turn the evolution equations for a \emph{diffusive variant of the Giesekus model}. Note that the system of evolution equations contains evolution equations for mechanical variables as well as for the \emph{temperature}.

If we set $\tilde{\mu}=0$, that is if the stress diffusion term vanishes, see below, then we obtain, for mechanical variables, the original model proposed by~\cite{giesekus.h:simple}. (No evolution equation for temperature was given in~\cite{giesekus.h:simple}.) In this sense the analysis outlined below also provides a thermodynamical background for the original Giesekus model.

\subsection{Helmholtz free energy}
\label{sec:giesekus-fenergy}
The \emph{ansatz} for the specific Helmholtz free energy for the Giesekus model is chosen as
\begin{equation}
  \label{eq:giesekus-fenergy}
  \fenergy 
  =_{\bydefinition} 
  \fenergy_0(\temp)
  + 
  \frac{\mu}{2\rho}
  \left[
    \Tr \lcgII
    -
    3
    -
    \ln \det \lcgII
  \right]
  ,
\end{equation}
where $\mu$ is a positive material constant and $\rho$ is a constant density. Notice that the Helmholtz free energy is, amongst others, specified in terms of the tensor~$\lcgII$, which is a special case of the general tensorial quantity $\lcgGSII$. We thus set $a = 1$, replace $\lcgGSII$ with $\lcgII$ in~\eqref{eq:entropy-evolution-from-fenergy}, and we consider the upper convected derivative $\fid{\generictensor}$ instead of the Gordon--Schowalter derivative $\gfid{\generictensor}$. 

The postulated Helmholtz free energy yields the following formulae for the partial derivatives of the Helmholtz free energy that appear in~\eqref{eq:entropy-evolution-from-fenergy}
\begin{subequations}
  \label{eq:giesekus-fenergy-pds}
  \begin{align}
  \pd{\fenergy}{\Tr \lcgII}
  &=
  \frac{\mu}{2 \rho},
  \\
  \pd{\fenergy}{\ln \det \lcgII}
  &=
  - \frac{\mu}{2 \rho},
  \\
  \pd{\fenergy}{\varphi}
  &=
  0.
  \end{align}
\end{subequations}
By plugging \eqref{eq:giesekus-fenergy-pds} into \eqref{eq:entropy-evolution-from-fenergy} we arrive at
\begin{equation}
  \label{eq:giesekus-entropy-evolution-from-fenergy}
  \rho \dd{\entropy}{t} 
  + 
  \divergence \left( \frac{\efluxc}{\temp} \right)
  =
  \frac{1}{\temp}
  \left\{
  \tensordot{
    \left[
    \traceless{\cstress} 
    - 
    \mu \traceless{\left( \lcgII \right)}
    \right]
  }{
    \traceless{\gradsym}
  }
  - \frac{\mu}{2} \Tr 
  \left[ 
  \fid{\overline{\lcgII}} \left( \identity - \inverse{\lcgII} \right) 
  \right]
  -
  \frac{\vectordot{\efluxc}{\nabla \temp}}{\temp}
  \right\}
  .
\end{equation}
This manipulation gives us the entropy evolution equation that is implied by the choice of the specific Helmholtz free energy.

\subsection{Entropy production}
\label{sec:giesekus-entropy-production}
Next, we postulate the entropy production for the Giesekus model with stress diffusion in the form
\begin{equation}
  \label{eq:giesekus-entropy-production}
  \entprodc
  =
  \frac{1}{\temp}
  \left\{
  2 \nu \tensordot{\traceless{\gradsym}}{\traceless{\gradsym}}
  +
  \frac{\mu^2}{2 \nu_1} \Tr 
  \left[ 
    \lambda \lcgII^2 + (1 - 3 \lambda) \lcgII + (1 + \lambda) \inverse{\lcgII} + (3 \lambda - 2) \identity
  \right]
  +
  \frac{\mu \tilde{\mu}(\temp)}{2\nu_1}
  \tensorddot{
    \nabla \lcgII
  }
  {
    \nabla \lcgII
  }
  +
  \kappa
  \vectordot{\nabla \temp}{\frac{\nabla \temp}{\temp}}
  \right\}
  ,
\end{equation}
where $\lambda \in [0,1]$ is a model parameter, $\nu$, $\nu_1$ are positive material constants, and $\tilde{\mu}$ is a material parameter that can be temperature dependent. The entropy production is chosen in such a way that it is a \emph{nonnegative function}, which automatically guarantees the \emph{fulfillment of the second law of thermodynamics}. The nonnegativity of the first, third and fourth term on the right-had side of~\eqref{eq:giesekus-entropy-production} is obvious. Concerning the third term, we know that~$\lcgII$ is a symmetric positive definite matrix, hence it is diagonalisable with the positive eigenvalues~$\left\{ \mu_i \right\}_{i=1}^3$~on the diagonal. Consequently, we see that
\begin{equation}
  \label{eq:53}
  \Tr
  \left[
    \lambda \lcgII^2 + (1 - 3 \lambda) \lcgII + (1 + \lambda) \inverse{\lcgII} + (3 \lambda - 2) \identity
  \right]
  =
  \sum_{i=1}^3
  \left(
    \lambda \mu_i^2 + (1 - 3 \lambda) \mu_i + (1 + \lambda) \frac{1}{\mu_i} + (3 \lambda - 2)
  \right)
  ,
\end{equation}
where the function $\lambda \mu_i^2 + (1 - 3 \lambda) \mu_i + (1 + \lambda) \frac{1}{\mu_i} + (3 \lambda - 2)$ is for $\lambda \in [0,1]$ and positive $\mu_i$ a nonnegative function that vanishes if and only if $\mu_i=1$. Finally, we can also observe that the entropy production vanishes in the equilibrium homogeneous rest state, that is if $\vec{v}=\vec{0}$, $\lcgII = \identity$, and $\temp = \temp_{\mathrm{eq}}$, where $\temp_{\mathrm{eq}}$ is a spatially uniform temperature field.

The general evolution equation for entropy
\begin{equation}
  \rho \dd{\entropy}{t} + \divergence \entfluxc = \entprodc,
\end{equation}
can then be transformed using the chosen entropy production $\entprodc$ and the identity \eqref{eq:tensorddot-identity}, where we set $c = \tilde{\mu}$, $\generictensor = \lcgII$, into the form
\begin{multline}
  \label{eq:giesekus-entropy-evolution-comparison}
  \rho \dd{\entropy}{t} 
  + 
  \divergence 
  \left\{ 
    \entfluxc 
    - 
    \frac{
      \frac{\mu \tilde{\mu}}{4 \nu_1} 
      \Tr
      \left[
        \left( \nabla \lcgII \right) \left( \lcgII - \identity \right)
        +
        \left( \lcgII - \identity \right) \left( \nabla \lcgII \right)
      \right]
    }{
      \temp
    }
  \right\}
  =
  \frac{1}{\temp}
  \Bigg\{
    2 \nu \tensordot{\traceless{\gradsym}}{\traceless{\gradsym}}
    \\
    +
    \frac{\mu}{2 \nu_1} \Tr 
    \left[ 
      \left[
        \mu
        \left[
          \lambda \lcgII^2 + (1 - 2 \lambda) \lcgII - (1 - \lambda) \identity
        \right]
        -
        \frac{1}{2}
        \left[
          \divergence \left( \tilde{\mu} \nabla \lcgII \right) \lcgII
          +
          \lcgII \divergence \left( \tilde{\mu} \nabla \lcgII \right)
        \right]
      \right]
      \left(
        \identity - \inverse{\lcgII}
      \right)
    \right]
    \\
    +
    \frac{
      \vectordot{
        \left[
          \kappa \nabla \temp
          +
          \frac{\mu \tilde{\mu}}{4 \nu_1} 
          \Tr
          \left[
            \left( \nabla \lcgII \right) \left( \lcgII - \identity \right)
            +
            \left( \lcgII - \identity \right) \left( \nabla \lcgII \right)
          \right]
        \right]
      }{
        \nabla \temp
      }
    }{
      \temp
    }
  \Bigg\}.
\end{multline}

\subsection{Constitutive relations}
\label{sec:giesekus-constitutive-relations}
Now we are in a position to read constitutive relations for $\cstress$, $\lcgII$, $\efluxc$ and $\entfluxc$ out of~\eqref{eq:giesekus-entropy-evolution-comparison} and~\eqref{eq:giesekus-entropy-evolution-from-fenergy}. As discussed previously, all that needs to be done is to compare~\eqref{eq:giesekus-entropy-evolution-comparison}, which is the evolution equation for the entropy that is a consequence of the choice of the entropy production, with~\eqref{eq:giesekus-entropy-evolution-from-fenergy}, which is the evolution equation for the entropy that is a consequence of the choice of the Helmholtz free energy. In particular, comparing the terms with the symmetric part of the velocity gradient, that is the term
\begin{subequations}
  \label{eq:12}
\begin{equation}
  \label{eq:5}
    2 \nu \tensordot{\traceless{\gradsym}}{\traceless{\gradsym}}
\end{equation}
in~\eqref{eq:giesekus-entropy-evolution-comparison} with the corresponding term
\begin{equation}
  \label{eq:6}
    \tensordot{
    \left[
    \traceless{\cstress} 
    - 
    \mu \traceless{\left( \lcgII \right)}
    \right]
  }{
    \traceless{\gradsym}
  }
\end{equation}
in~\eqref{eq:giesekus-entropy-evolution-from-fenergy}, we see that these terms are equal provided that the constitutive relation for the traceless part of the Cauchy stress tensor reads
\begin{equation}
  \label{eq:13}
  \traceless{\cstress}
  =
  2 \nu \traceless{\gradsym}
  +
  \mu \traceless{\left( \lcgII \right)},
  .  
\end{equation}
\end{subequations}

Making the comparison for the remaining terms in equations~\eqref{eq:giesekus-entropy-evolution-from-fenergy} and \eqref{eq:giesekus-entropy-evolution-comparison} then leads us to the constitutive relations
\begin{subequations}
  \label{eq:giesekus-constitutive-relations}
  \begin{align}
    \label{eq:giesekus-cstress}
    \traceless{\cstress}
    &=
    2 \nu \traceless{\gradsym}
    +
    \mu \traceless{\left( \lcgII \right)},
    \\
    \label{eq:giesekus-lcgII-evolution-equation}
    \nu_1 \fid{\overline{\lcgII}}
    &=
    -
    \mu
    \left[
    \lambda \lcgII^2 + (1 - 2 \lambda) \lcgII - (1 - \lambda) \identity
    \right]
    +
    \frac{1}{2}
    \left[
      \divergence \left( \tilde{\mu} \nabla \lcgII \right) \lcgII
      +
      \lcgII \divergence \left( \tilde{\mu} \nabla \lcgII \right)
    \right],
    \\
    \label{eq:giesekus-eflux}
    \efluxc
    &=
    -
    \kappa \nabla \temp
    -
    \frac{\mu \tilde{\mu}}{4 \nu_1}
    \Tr
    \left[
      \left( \nabla \lcgII \right) \left( \lcgII - \identity \right)
      +
      \left( \lcgII - \identity \right) \left( \nabla \lcgII \right)
    \right],
    \\
    \label{eq:giesekus-entflux}
    \entfluxc
    &=
    -\frac{\kappa \nabla \temp}{\temp}.
  \end{align}
\end{subequations}
Relations \eqref{eq:giesekus-constitutive-relations} are then used in the formulation of the evolution equations for the mechanical quantities as well as the temperature evolution equation. In particular, if we want to obtain the temperature evolution equation, we substitute for $\entprodc$ and $\entfluxc$ from \eqref{eq:giesekus-entropy-production} and \eqref{eq:giesekus-entflux} into the general temperature evolution equation \eqref{eq:temperature-evolution-general}.

\subsection{Governing equations}
\label{sec:giesekus-governing-equations}
Following the steps outlined above we see that the Helmholtz free energy ansatz~\eqref{eq:giesekus-fenergy} and the entropy production ansatz~\eqref{eq:giesekus-entropy-production} lead to the following full system of governing equations for the diffusive Giesekus model
\begin{subequations}
  \label{eq:giesekus-governing-equations}
  \begin{align}
    \label{eq:giesekus-incompressibility-condition}
    \divergence \vec{v} 
    &=
    0,
    \\
    \label{eq:giesekus-linear-momentum-balance}
    \rho \dd{\vec{v}}{t}
    &=
    \divergence \cstress,
    \\
    \label{eq:giesekus-lcg-evolution-equation}
    \nu_1 \fid{\overline{\lcgII}}
    &=
    -
    \mu
    \left[
      \lambda \lcgII^2 + (1 - 2 \lambda) \lcgII - (1 - \lambda) \identity
    \right]
    +
    \frac{1}{2}
    \left[
    \divergence \left( \tilde{\mu} \nabla \lcgII \right) \lcgII
    +
    \lcgII \divergence \left( \tilde{\mu} \nabla \lcgII \right)
    \right],
    \\
    \label{eq:giesekus-temperature-evolution-equation}
    \rho \cheatvol^{\mathrm{NSE}}
    \dd{\temp}{t}
    &=
    \divergence \left( \kappa \nabla \temp \right)
    +
    2 \nu \tensordot{\traceless{\gradsym}}{\traceless{\gradsym}}
    +
    \frac{\mu^2}{2 \nu_1} \Tr 
    \left[ 
      \lambda \lcgII^2 + (1 - 3 \lambda) \lcgII + (1 + \lambda) \inverse{\lcgII} + (3 \lambda - 2) \identity
    \right]
    +
    \frac{\mu \tilde{\mu}(\temp)}{2\nu_1}
    \tensorddot{
      \nabla \lcgII
    }
    {
      \nabla \lcgII
    },
  \end{align}
  where the Cauchy stress tensor $\cstress$ is given by the formulae
  \begin{equation}
    \label{eq:15}
    \cstress = \mns \identity + \traceless{\cstress},
    \qquad
    \traceless{\cstress} = 2 \nu \gradsym + \mu \traceless{(\lcgII)},
  \end{equation}
\end{subequations}
and where $-\mns$ denotes the pressure. (See~\cite{malek.j.prusa.v:derivation} and references therein concerning the notion of the pressure in incompressible fluids.) If body forces were present, the only change to the system would be the additional term $\rho \vec{b}$ in the balance of linear momentum~\eqref{eq:giesekus-linear-momentum-balance}, that is~\eqref{eq:giesekus-linear-momentum-balance} would read
\begin{equation}
  \label{eq:21}
  \rho \dd{\vec{v}}{t}
  =
  \divergence \cstress
  +
  \rho
  \vec{b}
  .
\end{equation}
For the sake of simplicity, we however ignore this term in the remaining analysis.

One can note that if $\tilde{\mu}=0$ and $\lambda = 0$, then the evolution equations for the mechanical variables, that is equations \eqref{eq:giesekus-incompressibility-condition}, \eqref{eq:giesekus-linear-momentum-balance} and \eqref{eq:giesekus-lcg-evolution-equation}, reduce to the governing equations of the standard Oldroyd-B fluid.

\section{FENE-P model}
\label{sec:fene-p-model}
Here, we identify the energy storage mechanisms and the entropy production mechanisms which yield the constitutive relations and in turn the evolution equations for a diffusive variant of the classical FENE-P, including the temperature evolution equation. The original model without the stress diffusion term was proposed by~\cite{bird.rb.dotson.pj.ea:polymer}, see also comments in~\cite{keunings.r:on}. The procedure for the derivation of the model is the same as in the case of the Giesekus model, hence we will proceed with less detail.

\subsection{Helmholtz free energy}
\label{sec:fene-p-fenergy}
The \emph{ansatz} for the specific Helmholtz free energy for the FENE-P model is chosen as
\begin{equation}
  \label{eq:fene-p-free-energy}
  \fenergy 
  =_{\bydefinition} 
  \fenergy_0(\temp)
  + 
  \frac{\mu}{2\rho}
  \left[
    -
    b \ln \left( 1 - \frac{1}{b} \Tr \lcgII \right) 
    -
    3
    -
    \ln \det \lcgII
  \right]
  .
\end{equation}
where $\mu$ is a positive material constant and $b$ is a positive model constant which represents an upper bound on the trace of the tensor $\lcgII$, that is  $\Tr \lcgII < b$. Notice that the Helmholtz free energy is, amongst others, specified in terms of the tensor $\lcgII$, the special case of the general tensorial quantity $\lcgGSII$. We thus set $a = 1$, replace $\lcgGSII$ with $\lcgII$ in \eqref{eq:entropy-evolution-from-fenergy}, and we again consider the upper convected derivative instead of the Gordon--Schowalter derivative. 

The postulated Helmholtz free energy yields the following formulae for the partial derivatives of the Helmholtz free energy that appear in \eqref{eq:entropy-evolution-from-fenergy}
\begin{subequations}
  \label{eq:fene-p-fenergy-pds}
  \begin{align}
    \pd{\fenergy}{\Tr \lcgII}
    &=
    \frac{\mu}{2 \rho} \inverse{\left( 1 - \frac{1}{b} \Tr \lcgII \right)},
    \\
    \pd{\fenergy}{\ln \det \lcgII}
    &=
    - \frac{\mu}{2 \rho},
    \\
    \pd{\fenergy}{\varphi}
    &=
    0.
  \end{align}
\end{subequations}
By plugging \eqref{eq:fene-p-fenergy-pds} into \eqref{eq:entropy-evolution-from-fenergy} we get
\begin{equation}
  \label{eq:fene-p-entropy-evolution-from-fenergy}
  \rho \dd{\entropy}{t} 
  + 
  \divergence \left( \frac{\efluxc}{\temp} \right)
  =
  \frac{1}{\temp}
  \left\{
  \tensordot{
    \left[
    \traceless{\cstress} 
    - 
    \mu \inverse{\left( 1 - \frac{1}{b} \Tr \lcgII \right)} \traceless{\left( \lcgII \right)}
    \right]
  }{
    \traceless{\gradsym}
  }
  - \frac{\mu}{2} \Tr 
  \left\{
  \fid{\overline{\lcgII}} 
  \left[ 
  \inverse{\left( 1 - \frac{1}{b} \Tr \lcgII \right)} \identity - \inverse{\lcgII} 
  \right] 
  \right\}
  -
  \frac{\vectordot{\efluxc}{\nabla \temp}}{\temp}
  \right\}.
\end{equation}

\subsection{Entropy production}
\label{sec:fene-p-entropy-production}
Now we postulate the entropy production for the FENE-P model with stress diffusion in the form
\begin{multline}
  \label{eq:fene-p-entropy-production}
  \entprodc
  =
  \frac{1}{\temp}
  \left\{
  2 \nu \tensordot{\traceless{\gradsym}}{\traceless{\gradsym}}
  +
  \frac{\mu^2}{2 \nu_1} 
  \left[ 
    \left( 1 - \frac{1}{b} \Tr \lcgII \right)^{-2} \Tr \lcgII
    -
    6 \inverse{\left( 1 - \frac{1}{b} \Tr \lcgII \right)}
    +
    \Tr \inverse{\lcgII}
  \right]
  \right.
  \\
  \left.
  +
  \frac{\mu \tilde{\mu}(\temp)}{2\nu_1}
  \tensorddot{
    \nabla 
    \left[ 
      \inverse{\left( 1 - \frac{1}{b} \Tr \lcgII \right)} \lcgII
    \right]
  }
  {
    \nabla 
    \left[ 
      \inverse{\left( 1 - \frac{1}{b} \Tr \lcgII \right)} \lcgII
    \right]
  }
  +
  \kappa
  \vectordot{\nabla \temp}{\frac{\nabla \temp}{\temp}}
  \right\},
\end{multline}
where $\nu$, $\nu_1$ are positive material constants, and $\tilde{\mu}$ is a (possibly) temperature-dependent material parameter. Since~$\lcgII$ is a symmetric positive definite matrix we can easily show that the second term of \eqref{eq:fene-p-entropy-production} is nonnegative by using the spectral decomposition of $\lcgII$, and by investigating the nonnegativity of the corresponding scalar function of eigenvalues of $\lcgII$. (See~\eqref{eq:53} for a detailed discussion concerning the same issue.) Consequently, the proposed entropy production is a nonnegative function. 

The general evolution equation for entropy
\begin{equation}
  \rho \dd{\entropy}{t} + \divergence \entfluxc = \entprodc,
\end{equation}
can then be transformed using the chosen entropy production $\entprodc$ and the identity \eqref{eq:tensorddot-identity}, where we set $c = \tilde{\mu}$, $\generictensor = \inverse{\left( 1 - \frac{1}{b} \Tr \lcgII \right)} \lcgII$, into the form
\begin{multline}
  \label{eq:fene-p-entropy-evolution-comparison}
  \rho \dd{\entropy}{t} 
  + 
  \divergence 
  \entfluxc
  \\
  - 
  \divergence 
  \left\{ 
  \frac{
    \frac{\mu \tilde{\mu}}{4 \nu_1}
    \Tr
    \left\{
      \left[ 
        \nabla
        \left( 
          \inverse{\left( 1 - \frac{1}{b} \Tr \lcgII \right)} \lcgII 
        \right)  
      \right]
      \left[ 
        \inverse{\left( 1 - \frac{1}{b} \Tr \lcgII \right)} \lcgII - \identity 
      \right]
      +
      \left[ 
        \inverse{\left( 1 - \frac{1}{b} \Tr \lcgII \right)} \lcgII - \identity 
      \right]
      \left[ 
        \nabla
        \left( 
          \inverse{\left( 1 - \frac{1}{b} \Tr \lcgII \right)} \lcgII 
        \right)  
      \right]    
    \right\}
  }{
    \temp
  }
  \right\}
  \\
  =
  \frac{1}{\temp}
  \Bigg\{
  2 \nu \tensordot{\traceless{\gradsym}}{\traceless{\gradsym}}
  \\
  +
  \frac{\mu}{2 \nu_1} \Tr 
  \left[ 
    \left[
      -
      \mu
      \left[
        \inverse{\left( 1 - \frac{1}{b} \Tr \lcgII \right)} \lcgII - \identity
      \right]
      +
      \frac{1}{2}
      \left\{
        \divergence 
        \left[ 
          \tilde{\mu} \nabla 
          \left( 
            \inverse{\left( 1 - \frac{1}{b} \Tr \lcgII \right)} \lcgII 
          \right) 
        \right] 
        \lcgII
        +
        \lcgII
        \divergence 
        \left[ 
          \tilde{\mu} \nabla 
          \left( 
            \inverse{\left( 1 - \frac{1}{b} \Tr \lcgII \right)} \lcgII 
          \right) 
        \right]
      \right\}
    \right]
  \right.
    \\
  \left.
    \left[
      \inverse{\left( 1 - \frac{1}{b} \Tr \lcgII \right)} \identity - \inverse{\lcgII}
    \right]
  \right]
  \\
  +
  \frac{
    \vectordot{
      \left[
        \frac{\mu \tilde{\mu}}{4 \nu_1}
        \Tr
        \left\{
          \left[ 
            \nabla
            \left( 
              \inverse{\left( 1 - \frac{1}{b} \Tr \lcgII \right)} \lcgII 
            \right)  
          \right]
          \left[ 
            \inverse{\left( 1 - \frac{1}{b} \Tr \lcgII \right)} \lcgII - \identity 
          \right]
          +
          \left[ 
            \inverse{\left( 1 - \frac{1}{b} \Tr \lcgII \right)} \lcgII - \identity 
          \right]
          \left[ 
            \nabla
            \left( 
              \inverse{\left( 1 - \frac{1}{b} \Tr \lcgII \right)} \lcgII 
            \right)  
          \right]    
        \right\}
      \right]
    }{
      \nabla \temp
    }
  }{
    \temp
  }
  \\
  +
    \frac{
    \vectordot{
      \kappa \nabla \temp
    }{
      \nabla \temp
    }
  }{
    \temp
  }
  \Bigg\}.
\end{multline}
Equation \eqref{eq:fene-p-entropy-evolution-comparison} can now be compared with the derived entropy evolution equation \eqref{eq:fene-p-entropy-evolution-from-fenergy}.

\subsection{Constitutive relations}
\label{sec:fene-p-constitutive-relations}
In particular, equations \eqref{eq:fene-p-entropy-evolution-from-fenergy} and \eqref{eq:fene-p-entropy-evolution-comparison} will coincide if we set
\begin{subequations}
  \label{eq:fene-p-constitutive-relations}
  \begin{equation}
    \label{eq:fene-p-cstress}
    \traceless{\cstress}
    =
    2 \nu \traceless{\gradsym}
    +
    \mu \inverse{\left( 1 - \frac{1}{b} \Tr \lcgII \right)} \traceless{\left( \lcgII \right)},
  \end{equation}
  \begin{equation}
    \label{eq:fene-p-lcgII-evolution-equation}
    \nu_1 \fid{\overline{\lcgII}}
    =
    -
    \mu
    \left[
    \inverse{\left( 1 - \frac{1}{b} \Tr \lcgII \right)} \lcgII - \identity
    \right]
    +
    \frac{1}{2}
    \left\{
      \divergence 
      \left[ 
        \tilde{\mu} \nabla 
        \left( 
          \inverse{\left( 1 - \frac{1}{b} \Tr \lcgII \right)} \lcgII 
        \right) 
      \right] 
      \lcgII
      +
      \lcgII
      \divergence 
      \left[ 
        \tilde{\mu} \nabla 
        \left( 
          \inverse{\left( 1 - \frac{1}{b} \Tr \lcgII \right)} \lcgII 
        \right) 
      \right]
    \right\},
  \end{equation}
  \begin{multline}
    \label{eq:fene-p-eflux}
    \efluxc
    =
    -
    \kappa \nabla \temp
    -
    \frac{\mu \tilde{\mu}}{4 \nu_1}
    \Tr
    \left\{
      \left[ 
        \nabla
        \left( 
          \inverse{\left( 1 - \frac{1}{b} \Tr \lcgII \right)} \lcgII 
        \right)  
      \right]
      \left[ 
        \inverse{\left( 1 - \frac{1}{b} \Tr \lcgII \right)} \lcgII - \identity 
      \right]
    \right.
    \\
    \left.
      +
      \left[ 
        \inverse{\left( 1 - \frac{1}{b} \Tr \lcgII \right)} \lcgII - \identity 
      \right]
      \left[ 
        \nabla
        \left( 
          \inverse{\left( 1 - \frac{1}{b} \Tr \lcgII \right)} \lcgII 
        \right)  
      \right]    
    \right\},
  \end{multline}
  \begin{equation}
    \label{eq:fene-p-entflux}
    \entfluxc
    =
    -\frac{\kappa \nabla \temp}{\temp}
    .
  \end{equation}
\end{subequations}
Relations \eqref{eq:fene-p-constitutive-relations} can then be used for formulating the evolution equations for the mechanical quantities as well as the temperature evolution equation.

\subsection{Governing equations}
\label{sec:fene-p-governing-equations}
Following the steps outlined above we see that the Helmholtz free energy ansatz~\eqref{eq:fene-p-free-energy} and the entropy production ansatz~\eqref{eq:fene-p-entropy-production} lead to the following full system of governing equations for the diffusive FENE-P model
\begin{subequations}
  \label{eq:fene-p-governing-equations}
  \begin{align}
    \label{eq:fene-p-incompressibility-condition}
    \divergence \vec{v} 
    &=
    0,
    \\
    \label{eq:fene-p-linear-momentum-balance}
    \rho \dd{\vec{v}}{t}
    &=
    \divergence \cstress,
    \\
    \label{eq:fene-p-lcg-evolution-equation}
    \nu_1 \fid{\overline{\lcgII}}
    &=
    -
    \mu
    \left[
    \inverse{\left( 1 - \frac{1}{b} \Tr \lcgII \right)} \lcgII - \identity
    \right]
    +
    \frac{1}{2}
    \left\{
      \divergence 
      \left[ 
        \tilde{\mu} \nabla 
        \left( 
          \inverse{\left( 1 - \frac{1}{b} \Tr \lcgII \right)} \lcgII 
        \right) 
      \right] 
      \lcgII
      +
      \lcgII
      \divergence 
      \left[ 
        \tilde{\mu} \nabla 
        \left( 
          \inverse{\left( 1 - \frac{1}{b} \Tr \lcgII \right)} \lcgII 
        \right) 
      \right]
    \right\},
  \end{align}
  \begin{multline}
    \label{eq:fene-p-temperature-evolution-equation}
    \rho \cheatvol^{\mathrm{NSE}}
    \dd{\temp}{t}
    =
    \divergence \left( \kappa \nabla \temp \right)
    +
    2 \nu \tensordot{\traceless{\gradsym}}{\traceless{\gradsym}}
    +
    \frac{\mu^2}{2 \nu_1} 
    \left[ 
    \left( 1 - \frac{1}{b} \Tr \lcgII \right)^{-2} \Tr \lcgII
    -
    6 \inverse{\left( 1 - \frac{1}{b} \Tr \lcgII \right)}
    +
    \Tr \inverse{\lcgII}
    \right]
    \\
    +
    \frac{\mu \tilde{\mu}(\temp)}{2\nu_1}
    \tensorddot{
      \nabla 
      \left[ 
      \inverse{\left( 1 - \frac{1}{b} \Tr \lcgII \right)} \lcgII
      \right]
    }
    {
      \nabla 
      \left[ 
      \inverse{\left( 1 - \frac{1}{b} \Tr \lcgII \right)} \lcgII
      \right]
    },
  \end{multline}
  where the Cauchy stress tensor $\cstress$ is given by the formulae
  \begin{equation}
    \cstress = \mns \identity + \traceless{\cstress},
    \qquad
    \traceless{\cstress} = 2 \nu \gradsym + \mu \inverse{\left( 1 - \frac{1}{b} \Tr \lcgII \right)} \traceless{(\lcgII)}.
  \end{equation}
\end{subequations}
We again see that if $\tilde{\mu} = 0$, that is if the stress diffusion term vanishes, then the governing equations for the mechanical variables coincide with the governing equations for the original FENE-P model without stress diffusion.

\section{Johnson--Segalman model}
\label{sec:js-model}
Here, we identify the energy storage mechanisms and the entropy production mechanisms which yield the constitutive relations and in turn the evolution equations for a diffusive variant of the Johnson--Segalman model, including the temperature evolution equation. The original model without the stress diffusion term was proposed by~\cite{johnson.mw.segalman.d:model}. Unlike in the previous sections devoted to the analysis of the Giesekus and FENE-P models, we need to change the kinematical assumptions, since we want to obtain a model with the Gordon--Schowalter derivative instead of the upper convected derivative. The kinematical assumptions that lead to the Gordon--Schowalter derivative have been already discussed in the introductory sections, hence, regarding the discussion of the kinematics, we restrict ourselves to a simple statement that we are going to replace tensor $\lcgII$ by $\lcgGSII$.

\subsection{Helmholtz free energy}
\label{sec:js-fenergy}
The \emph{ansatz} for the Helmholtz free energy of the Johnson--Segalman model is chosen as
\begin{equation}
  \label{eq:js-free-energy}
  \fenergy 
  =_{\bydefinition} 
  \fenergy_0(\temp)
  + 
  \frac{\mu}{2\rho}
  \left(
  \Tr \lcgGSII
  -
  3
  -
  \ln \det \lcgGSII
  \right)
  ,
\end{equation}
where $\mu$ is a positive material constant. Note that the chosen formula for the Helmholtz free energy is structurally identical to that for the Giesekus model, the only difference between~\eqref{eq:js-free-energy} and~\eqref{eq:giesekus-fenergy} is that $\lcgII$ has been replaced by~$\lcgGSII$. The postulated Helmholtz free energy yields the following formulae for the partial derivatives that appear in \eqref{eq:entropy-evolution-from-fenergy}
\begin{subequations}
  \label{eq:js-fenergy-pds}
  \begin{align}
  \pd{\fenergy}{\Tr \lcgGSII}
  &=
  \frac{\mu}{2 \rho},
  \\
  \pd{\fenergy}{\ln \det \lcgGSII}
  &=
  - \frac{\mu}{2 \rho},
  \\
  \pd{\fenergy}{\varphi}
  &=
  0.
  \end{align}
\end{subequations}
By plugging \eqref{eq:js-fenergy-pds} into \eqref{eq:entropy-evolution-from-fenergy} we get
\begin{equation}
  \label{eq:js-entropy-evolution-from-fenergy}
  \rho \dd{\entropy}{t} 
  + 
  \divergence \left( \frac{\efluxc}{\temp} \right)
  =
  \frac{1}{\temp}
  \left\{
  \tensordot{
    \left[
    \traceless{\cstress}
    -
    a \mu \traceless{\left( \lcgGSII \right)}
    \right]
  }{
    \traceless{\gradsym}
  }
  - 
  \frac{\mu}{2}
  \Tr
  \left[
    \gfid{\overline{\lcgGSII}}
    \left(
      \identity
      -
      \inverse{\lcgGSII}
    \right)
  \right]
  -
  \frac{\vectordot{\efluxc}{\nabla \temp}}{\temp}
  \right\}.
\end{equation}

\subsection{Entropy production}
\label{sec:js-entropy-production}
We postulate the entropy production for the Johnson--Segalman model with stress diffusion in the form
\begin{equation}
  \label{eq:js-entropy-production}
  \entprodc
  =
  \frac{1}{\temp}
  \left\{
  2 \nu \tensordot{\traceless{\gradsym}}{\traceless{\gradsym}}
  +
  \frac{\mu^2}{2 \nu_1} 
  \left[ 
    \Tr \lcgGSII + \Tr \inverse{\lcgGSII} - 6
  \right]
  +
  \frac{\mu \tilde{\mu}(\temp)}{2\nu_1}
  \tensorddot{
    \nabla \lcgGSII
  }
  {
    \nabla \lcgGSII
  }
  +
  \kappa
  \vectordot{\nabla \temp}{\frac{\nabla \temp}{\temp}}
  \right\},
\end{equation}
where $\nu$, $\nu_1$ are positive material constants, and $\tilde{\mu}$ is a (possibly) temperature-dependent material parameter. The chosen entropy production is a nonnegative function. The general evolution equation for entropy
\begin{equation}
\rho \dd{\entropy}{t} + \divergence \entfluxc = \entprodc,
\end{equation}
can then be transformed using the chosen entropy production $\entprodc$ and the identity \eqref{eq:tensorddot-identity}, where we set $c = \tilde{\mu}$, $\generictensor = \lcgGSII$, into the form
\begin{multline}
  \label{eq:js-entropy-evolution-comparison}
  \rho \dd{\entropy}{t} 
  + 
  \divergence 
  \left\{ 
  \entfluxc 
  - 
  \frac{
    \frac{\mu \tilde{\mu}}{4 \nu_1} 
    \Tr
    \left[
      \left( \nabla \lcgGSII \right) \left( \lcgGSII - \identity \right)
      +
      \left( \lcgGSII - \identity \right) \left( \nabla \lcgGSII \right)
    \right]
  }{
    \temp
  }
  \right\}
  =
  \frac{1}{\temp}
  \bigg\{
  2 \nu \tensordot{\traceless{\gradsym}}{\traceless{\gradsym}}
  \\
  +
  \frac{\mu}{2 \nu_1} \Tr 
  \left[ 
    \left[
      \mu
      \left(
        \lcgGSII - \identity
      \right)
      -
      \frac{1}{2}
      \left[
        \divergence \left( \tilde{\mu} \nabla \lcgGSII \right) \lcgGSII
        +
        \lcgGSII \divergence \left( \tilde{\mu} \nabla \lcgGSII \right)
      \right]
    \right]
    \left(
    \identity - \inverse{\lcgGSII}
    \right)
  \right]
  \\
  +
  \frac{
    \vectordot{
      \left[
        \kappa \nabla \temp
        +
        \frac{\mu \tilde{\mu}}{4 \nu_1} 
        \Tr
        \left[
          \left( \nabla \lcgGSII \right) \left( \lcgGSII - \identity \right)
          +
          \left( \lcgGSII - \identity \right) \left( \nabla \lcgGSII \right)
        \right]
      \right]
    }{
      \nabla \temp
    }
  }{
    \temp
  }
  \bigg\}
\end{multline}
Equation \eqref{eq:js-entropy-evolution-comparison} can now be compared with the derived entropy evolution equation \eqref{eq:js-entropy-evolution-from-fenergy} to yield constitutive relations.

\subsection{Constitutive relations}
\label{sec:js-constitutive-relations}
In particular, equations \eqref{eq:js-entropy-evolution-from-fenergy} and \eqref{eq:js-entropy-evolution-comparison} will coincide if we set
\begin{subequations}
  \label{eq:js-constitutive-relations}
  \begin{align}
  \label{eq:js-cstress}
  \traceless{\cstress}
  &=
  2 \nu \traceless{\gradsym}
  +
  a \mu \traceless{\left( \lcgGSII \right)},
  \\
  \label{eq:js-lcgGSII-evolution-equation}
  \nu_1 \gfid{\overline{\lcgGSII}}
  &=
  -
  \mu
  \left(
    \lcgGSII - \identity
  \right)
  +
  \frac{1}{2}
  \left[
  \divergence \left( \tilde{\mu} \nabla \lcgGSII \right) \lcgGSII
  +
  \lcgGSII \divergence \left( \tilde{\mu} \nabla \lcgGSII \right)
  \right],
  \\
  \label{eq:js-eflux}
  \efluxc
  &=
  -
  \kappa \nabla \temp
  -
  \frac{\mu \tilde{\mu}}{4 \nu_1}
  \Tr
  \left[
  \left( \nabla \lcgGSII \right) \left( \lcgGSII - \identity \right)
  +
  \left( \lcgGSII - \identity \right) \left( \nabla \lcgGSII \right)
  \right],
  \\
  \label{eq:js-entflux}
  \entfluxc
  &=
  -\frac{\kappa \nabla \temp}{\temp}
  .
  \end{align}
\end{subequations}
Relations \eqref{eq:js-constitutive-relations} can then be used for formulating the evolution equations for the mechanical quantities as well as the temperature evolution equation.

\subsection{Governing equations}
\label{sec:js-governing-equations}
The full system of governing equations for the Johnson-Segalman model with stress diffusion reads
\begin{subequations}
  \label{eq:js-governing-equations}
  \begin{align}
    \label{eq:js-incompressibility-condition}
    \divergence \vec{v} 
    &=
    0,
    \\
    \label{eq:js-linear-momentum-balance}
    \rho \dd{\vec{v}}{t}
    &=
    \divergence \cstress,
    \\
    \label{eq:js-lcg-evolution-equation}
    \nu_1 \gfid{\overline{\lcgGSII}}
    &=
    -
    \mu
    \left(
      \lcgGSII - \identity
    \right)
    +
    \frac{1}{2}
    \left[
    \divergence \left( \tilde{\mu} \nabla \lcgGSII \right) \lcgGSII
    +
    \lcgGSII \divergence \left( \tilde{\mu} \nabla \lcgGSII \right)
    \right],
    \\
    \label{eq:js-temperature-evolution-equation}
    \rho \cheatvol^{\mathrm{NSE}}
    \dd{\temp}{t}
    &=
    \divergence \left( \kappa \nabla \temp \right)
    +
    2 \nu \tensordot{\traceless{\gradsym}}{\traceless{\gradsym}}
    +
    \frac{\mu^2}{2 \nu_1} 
    \left[ 
      \Tr \lcgGSII + \Tr \inverse{\lcgGSII} - 6
    \right]
    +
    \frac{\mu \tilde{\mu}(\temp)}{2\nu_1}
    \tensorddot{
      \nabla \lcgGSII
    }
    {
      \nabla \lcgGSII
    },
  \end{align}
  where the Cauchy stress tensor $\cstress$ is given by the formulae
  \begin{equation}
    \label{eq:14}
    \cstress = \mns \identity + \traceless{\cstress},
    \qquad
    \traceless{\cstress} = 2 \nu \gradsym + a \mu \traceless{(\lcgGSII)},
  \end{equation}
\end{subequations}
We note that the governing equations are nearly the same as the governing equations for the diffusive Oldroyd-B model. (Recall that Oldroyd-B model is the Giesekus model with $\lambda=0$, see~\eqref{eq:giesekus-governing-equations}.) This is not surprising given the fact that the formulae for the Helmholtz free energy and the entropy production are structurally the same. The difference between Oldroyd-B model and the Johnson--Segalman models is in the choice of the kinematical quantity $\lcgGSII$ versus $\lcgII$. This different choice is reflected in the final governing equations via the presence of the Gordon--Schowalter/upper convected derivative in the evolution equations, and by the presence of the factor $a$ in the formula for the Cauchy stress tensor, see~\eqref{eq:14} and~\eqref{eq:15}.

\section{Phan--Thien--Tanner model}
\label{sec:ptt-model}
Here, we identify the energy storage mechanisms and the entropy production mechanisms which yield the constitutive relations and in turn the evolution equations for a diffusive variant of the Phan--Thien--Tanner model, including the temperature evolution equation. The Phan--Thien--Tanner model without a stress diffusion term has several variants. The original model was proposed by~\cite{phan-thien.n.tanner.ri:new} and it corresponds to case \eqref{eq:linear-ptt} for the choice of the function appearing in the evolution equation for the extra stress tensor, see below. The exponential variant \eqref{eq:exponential-ptt} was proposed by~\cite{phan-thien.n:non-linear}. The quadratic variant \eqref{eq:quadratic-ptt} is far less seen in the literature, but is used for example by~\cite{ngamaramvaranggul.v.webster.m.f.:simulation}.

\subsection{Helmholtz free energy}
\label{sec:ptt-fenergy}
The \emph{ansatz} for the Helmholtz free energy of the Phan--Thien--Tanner model is chosen as
\begin{equation}
  \label{eq:ptt-free-energy}
  \fenergy 
  =_{\bydefinition} 
  \fenergy_0(\temp)
  + 
  \frac{\mu}{2\rho}
  \left[
    \Tr \lcgGSII
    -
    3
    -
    \ln \det \lcgGSII
  \right]
  ,
\end{equation}
where $\mu$ is a positive material constant. The postulated Helmholtz free energy yields the following formulae for the partial derivatives that appear in \eqref{eq:entropy-evolution-from-fenergy}
\begin{subequations}
  \label{eq:ptt-fenergy-pds}
  \begin{align}
    \pd{\fenergy}{\Tr \lcgGSII}
    &=
    \frac{\mu}{2 \rho},
    \\
    \pd{\fenergy}{\ln \det \lcgGSII}
    &=
    - \frac{\mu}{2 \rho},
    \\
    \pd{\fenergy}{\varphi}
    &=
    0.
  \end{align}
\end{subequations}
By plugging \eqref{eq:ptt-fenergy-pds} into \eqref{eq:entropy-evolution-from-fenergy} we get
\begin{equation}
  \label{eq:ptt-entropy-evolution-from-fenergy}
  \rho \dd{\entropy}{t} 
  + 
  \divergence \left( \frac{\efluxc}{\temp} \right)
  =
  \frac{1}{\temp}
  \left\{
  \tensordot{
    \left[
    \traceless{\cstress}
    -
    a \mu \traceless{\left( \lcgGSII \right)}
    \right]
  }{
    \traceless{\gradsym}
  }
  - 
  \frac{\mu}{2}
  \Tr
  \left[
    \gfid{\overline{\lcgGSII}}
    \left(
      \identity
      -
      \inverse{\lcgGSII}
    \right)
  \right]
  -
  \frac{\vectordot{\efluxc}{\nabla \temp}}{\temp}
  \right\}.
\end{equation}

\subsection{Entropy production}
\label{sec:ptt-entropy-production}
We postulate the entropy production for the Phan--Thien--Tanner model with stress diffusion in the form
\begin{equation}
  \label{eq:ptt-entropy-production}
  \entprodc
  =
  \frac{1}{\temp}
  \left\{
  2 \nu \tensordot{\traceless{\gradsym}}{\traceless{\gradsym}}
  +
  \frac{\mu^2}{2 \nu_1} f(\lcgGSII)
  \left[ 
    \Tr \lcgGSII + \Tr \inverse{\lcgGSII} - 6
  \right]
  +
  \frac{\mu \tilde{\mu}(\temp)}{2\nu_1}
  \tensorddot{
    \nabla \lcgGSII
  }
  {
    \nabla \lcgGSII
  }
  +
  \kappa
  \vectordot{\nabla \temp}{\frac{\nabla \temp}{\temp}}
  \right\},
\end{equation}
where $\nu$, $\nu_1$ are the positive material constants, $\tilde{\mu}$ is a (possibly) temperature-dependent material parameter, and the function $f$ is defined by one of the formulae
\begin{subnumcases}{f(\generictensor)=}
  \label{eq:linear-ptt}
  1 + p \Tr 
  \left( 
    \generictensor - \identity 
  \right),
  \\
  \label{eq:quadratic-ptt}
  1 + p \Tr 
  \left( 
    \generictensor - \identity 
  \right) 
  + 
  \left[ 
    p \Tr \left( \generictensor - \identity \right) 
  \right]^2,
  \\
  \label{eq:exponential-ptt}
  \exp 
  \left[
    p \Tr \left( \generictensor - \identity \right) 
  \right],
\end{subnumcases}
where $p$ is a model parameter. Since the function $f$ appears in the entropy production~\eqref{eq:ptt-entropy-production}, we see that the nonnegativity of the entropy production will be guaranteed provided that we will place some restrictions on $f$.

In particular, we want $f$ to be nonnegative for any symmetric positive definite matrix. The nonnegativity of $f$ is guaranteed for any $p$ provided that $f$ is given by~\eqref{eq:exponential-ptt}. On the other hand, if $f$ is given by~\eqref{eq:linear-ptt}, then we need to restrict the values of $p$ to the interval $\left[ 0 ,\frac{1}{3} \right]$. This requirement follows from the spectral decomposition. If $\generictensor$ is a symmetric positive definite matrix, then
\begin{equation*}
 1 + p \Tr \left( \generictensor - \identity \right) = 1 + p \left(\lambda_1 + \lambda_2 + \lambda_3\right) - 3p, 
\end{equation*}
where $\left\{ \lambda_i \right\}_{i=1}^3$ are positive eigenvalues of $\generictensor$. Consequently $1 + p \Tr \left( \generictensor - \identity \right) > 0$ provided that $p \in \left[ 0 ,\frac{1}{3} \right]$. Similarly, if $f$ is given by~\eqref{eq:quadratic-ptt}, then we again need to restrict $p$ to the interval $\left[ 0 ,\frac{1}{3} \right]$.

Having the \emph{ansatz} for the entropy production $\entprodc$, we see that the general evolution equation for entropy
\begin{equation}
\rho \dd{\entropy}{t} + \divergence \entfluxc = \entprodc,
\end{equation}
can then be transformed using the chosen entropy production $\entprodc$ and the identity \eqref{eq:tensorddot-identity}, where we set $c = \tilde{\mu}$, $\generictensor = \lcgGSII$, into the following form
\begin{multline}
  \label{eq:ptt-entropy-evolution-comparison}
  \rho \dd{\entropy}{t} 
  + 
  \divergence 
  \left\{ 
    \entfluxc 
    - 
    \frac{
      \frac{\mu \tilde{\mu}}{4 \nu_1} 
      \Tr
      \left[
        \left( \nabla \lcgGSII \right) \left( \lcgGSII - \identity \right)
        +
        \left( \lcgGSII - \identity \right) \left( \nabla \lcgGSII \right)
      \right]
    }{
      \temp
    }
  \right\}
  =
  \frac{1}{\temp}
  \Bigg\{
    2 \nu \tensordot{\traceless{\gradsym}}{\traceless{\gradsym}}
    \\
    +
    \frac{\mu}{2 \nu_1} \Tr 
    \left[ 
      \left[
        \mu f(\lcgGSII) 
        \left(
          \lcgGSII - \identity
        \right)
        -
        \frac{1}{2}
        \left[
          \divergence \left( \tilde{\mu} \nabla \lcgGSII \right) \lcgGSII
          +
          \lcgGSII \divergence \left( \tilde{\mu} \nabla \lcgGSII \right)
        \right]
      \right]
      \left(
        \identity - \inverse{\lcgGSII}
      \right)
    \right]
    \\
    +
    \frac{
      \vectordot{
        \left[
          \kappa \nabla \temp
          +
          \frac{\mu \tilde{\mu}}{4 \nu_1} 
          \Tr
          \left[
            \left( \nabla \lcgGSII \right) \left( \lcgGSII - \identity \right)
            +
            \left( \lcgGSII - \identity \right) \left( \nabla \lcgGSII \right)
          \right]
        \right]
      }{
        \nabla \temp
      }
    }{
      \temp
    }
  \Bigg\}.
\end{multline}
Equation \eqref{eq:ptt-entropy-evolution-comparison} can now be compared with the derived entropy evolution equation \eqref{eq:ptt-entropy-evolution-from-fenergy} to yield constitutive relations.

\subsection{Constitutive relations}
\label{sec:ptt-constitutive-relations}
In particular, equations \eqref{eq:ptt-entropy-evolution-from-fenergy} and \eqref{eq:ptt-entropy-evolution-comparison} will coincide if we set
\begin{subequations}
  \label{eq:ptt-constitutive-relations}
  \begin{align}
  \label{eq:ptt-cstress}
  \traceless{\cstress}
  &=
  2 \nu \traceless{\gradsym}
  +
  a \mu \traceless{\left( \lcgGSII \right)},
  \\
  \label{eq:ptt-lcgGSII-evolution-equation}
  \nu_1 \gfid{\overline{\lcgGSII}}
  &=
  -
  \mu f(\lcgGSII)
  \left(
  \lcgGSII - \identity
  \right)
  +
  \frac{1}{2}
  \left[
  \divergence \left( \tilde{\mu} \nabla \lcgGSII \right) \lcgGSII
  +
  \lcgGSII \divergence \left( \tilde{\mu} \nabla \lcgGSII \right)
  \right],
  \\
  \label{eq:ptt-eflux}
  \efluxc
  &=
  -
  \kappa \nabla \temp
  -
  \frac{\mu \tilde{\mu}}{4 \nu_1}
  \Tr
  \left[
  \left( \nabla \lcgGSII \right) \left( \lcgGSII - \identity \right)
  +
  \left( \lcgGSII - \identity \right) \left( \nabla \lcgGSII \right)
  \right],
  \\
  \label{eq:ptt-entflux}
  \entfluxc
  &=
  -\frac{\kappa \nabla \temp}{\temp}
  .
  \end{align}
\end{subequations}
Relations \eqref{eq:ptt-constitutive-relations} can then be used for formulating the evolution equations for the mechanical quantities as well as the temperature evolution equation.

\subsection{Governing equations}
\label{sec:ptt-governing-equations}
The full system of governing equations for the diffusive variant of the Phan-Thien--Tanner model reads
\begin{subequations}
  \label{eq:ptt-governing-equations}
  \begin{align}
    \label{eq:ptt-incompressibility-condition}
    \divergence \vec{v} 
    &=
    0,
    \\
    \label{eq:ptt-linear-momentum-balance}
    \rho \dd{\vec{v}}{t}
    &=
    \divergence \cstress,
    \\
    \label{eq:ptt-lcg-evolution-equation}
    \nu_1 \gfid{\overline{\lcgGSII}}
    &=
    -
    \mu f(\lcgGSII)
    \left(
      \lcgGSII - \identity
    \right)
    +
    \frac{1}{2}
    \left[
    \divergence \left( \tilde{\mu} \nabla \lcgGSII \right) \lcgGSII
    +
    \lcgGSII \divergence \left( \tilde{\mu} \nabla \lcgGSII \right)
    \right],
    \\
    \label{eq:ptt-temperature-evolution-equation}
    \rho \cheatvol^{\mathrm{NSE}}
    \dd{\temp}{t}
    &=
    \divergence \left( \kappa \nabla \temp \right)
    +
    2 \nu \tensordot{\traceless{\gradsym}}{\traceless{\gradsym}}
    +
    \frac{\mu^2}{2 \nu_1} f(\lcgGSII)
    \left[ 
      \Tr \lcgGSII + \Tr \inverse{\lcgGSII} - 6
    \right]
    +
    \frac{\mu \tilde{\mu}(\temp)}{2\nu_1}
    \tensorddot{
      \nabla \lcgGSII
    }
    {
      \nabla \lcgGSII
    },
  \end{align}
  where the Cauchy stress tensor $\cstress$ is given by the formulae
  \begin{equation}
    \cstress = \mns \identity + \traceless{\cstress},
    \qquad
    \traceless{\cstress} = 2 \nu \gradsym + a \mu \traceless{(\lcgGSII)},
  \end{equation}
\end{subequations}
and the function $f$ is given by \eqref{eq:linear-ptt}, \eqref{eq:quadratic-ptt} or~\eqref{eq:exponential-ptt}. Adjusting the value of the parametr $a$ we can again select the convenient frame-indifferent derivative in the evolution equation for the extra stress tensor. Note also that if we have set $f=1$, then we would obtain the Johnson--Segalman model.

\section{Bautista--Manero--Puig type model}
\label{sec:bmp-model}
Here, we identify the energy storage mechanisms and the entropy production mechanisms which yield the constitutive relations and in turn the evolution equations for a diffusive variant of Bautista--Manero--Puig model, including the temperature evolution equation.

The Bautista--Manero-Puig type models are based on the idea that the fluid can undergo microstructural changes that are triggered by the flow. In this respect, this class of models is a generalisation of the original idea by~\cite{fredrickson.ag:model}. \cite{fredrickson.ag:model} has considered the standard Navier--Stokes fluid, but he has assumed that the fluidity (the reciprocal value of viscosity) is given by an extra evolution equation. Bautista--Manero--Puig type models basically extend this idea to viscoelastic rate-type models, see~\cite{bautista.f.santos.jm.ea:understanding} and the follow-up studies by~\cite{bautista.f.soltero.jfa.ea:on,bautista.f.soltero.jfa.ea:irreversible,manero.o.perez-lopez.jh.ea:thermodynamic,lopez-aguilar.je.webster.mf.ea:comparative}.

The model presented below is based on the same idea that leads to the Bautista--Manero--Puig model. However, even if we consider the model without the stress diffusion term, then the present model is in certain aspects \emph{different} from the original Bautista--Manero--Puig model. We will discuss this issue when we come to the point where the difference arises.

\subsection{Helmholtz free energy}
The \emph{ansatz} for the Helmholtz free energy for the diffusive Bautista--Manero--Puig type model is chosen as
\label{sec:bmp-fenergy}
\begin{equation}
  \label{eq:bmp-fenergy}
  \fenergy 
  =_{\bydefinition} 
  \fenergy_0(\temp)
  + 
  \frac{\mu}{2\rho}
  \left[
    \Tr \lcgII
    -
    3
    -
    \ln \det \lcgII
  \right]
  +
  \frac{\chi}{2 \rho}
  \left(
    \varphi_0 - \varphi
  \right)^2
  .
\end{equation}
where $\mu$, $\chi$, and $\varphi_0$ are positive material parameters. Notice that the Helmholtz free energy is a function of the tensor~$\lcgII$, which is a special instance of the general tensorial quantity $\lcgGSII$. We thus set $a = 1$, and we replace~$\lcgGSII$ with~$\lcgII$ in \eqref{eq:entropy-evolution-from-fenergy}. This will lead us to the use of the upper convected derivative instead of the Gordon--Schowalter derivative in the corresponding governing equation. (See also the derivation of the Giesekus model for a similar discussion.) The major difference between the Helmholtz free energy \emph{ansatz} used in the previous sections and the Helmholtz free energy \emph{ansatz}~\eqref{eq:bmp-fenergy} used in the current section is the presence of a new positive scalar variable $\varphi$. This variable serves as a proxy to account for microstructural changes.

The postulated Helmholtz free energy yields the following formulae for the partial derivatives that appear in~\eqref{eq:entropy-evolution-from-fenergy}
\begin{subequations}
  \label{eq:bmp-fenergy-pds}
  \begin{align}
    \pd{\fenergy}{\Tr \lcgII}
    &=
    \frac{\mu}{2 \rho},
    \\
    \pd{\fenergy}{\ln \det \lcgII}
    &=
    - \frac{\mu}{2 \rho},
    \\
    \pd{\fenergy}{\varphi}
    &=
      -
      \frac{\chi}{\rho}
      \left(\varphi_0 - \varphi \right).
  \end{align}
\end{subequations}
By plugging \eqref{eq:bmp-fenergy-pds} into \eqref{eq:entropy-evolution-from-fenergy} we get
\begin{equation}
  \label{eq:bmp-entropy-evolution-from-fenergy}
  \rho \dd{\entropy}{t} 
  + 
  \divergence \left( \frac{\efluxc}{\temp} \right)
  =
  \frac{1}{\temp}
  \left\{
  \tensordot{
    \left[
    \traceless{\cstress}
    -
    \mu \traceless{\left( \lcgII \right)}
    \right]
  }{
    \traceless{\gradsym}
  }
  - 
  \frac{\mu}{2}
  \Tr
  \left[
  \fid{\overline{\lcgII}}
    \left(
    \identity
    -
    \inverse{\lcgII}
    \right)
  \right]
  +
  \chi
  \left( \varphi_0 - \varphi \right) \dd{\varphi}{t}
  -
  \frac{\vectordot{\efluxc}{\nabla \temp}}{\temp}
  \right\}.
\end{equation}

\subsection{Entropy production}
\label{sec:bmp-entropy-production}
The entropy production for the Bautista--Manero--Puig type model with stress diffusion is postulated in the form
\begin{equation}
  \label{eq:bmp-entropy-production}
  \entprodc
  =
  \frac{1}{\temp}
  \left\{
  2 \nu \tensordot{\traceless{\gradsym}}{\traceless{\gradsym}}
  +
  \frac{\mu^2 \varphi}{2} 
  \left[ 
    \Tr \lcgII + \Tr \inverse{\lcgII} - 6
  \right]
  +
  \frac{\beta \mu \tilde{\mu}}{2}
  \tensorddot{
    \nabla \lcgII
  }
  {
    \nabla \lcgII
  }
  +
  \chi
  \frac{\left( \varphi_0 - \varphi \right)^2}{\tau}
  +
  \kappa
  \vectordot{\nabla \temp}{\frac{\nabla \temp}{\temp}}
  \right\},
\end{equation}
where $\nu$ and $\beta$ are positive material constants, and $\tilde{\mu}$ is a (possibly) temperature-dependent material parameter. \emph{We note that the chosen entropy production is nonnegative provided that we can guarantee that $\varphi$ is a positive quantity}, which is the issue we will discuss later. The general evolution equation for entropy
\begin{equation}
\rho \dd{\entropy}{t} + \divergence \entfluxc = \entprodc,
\end{equation}
can then be transformed using the chosen entropy production $\entprodc$ and the identity \eqref{eq:tensorddot-identity}, where we set $c = \tilde{\mu}$, $\generictensor = \lcgII$, into the form
\begin{multline}
  \label{eq:bmp-entropy-evolution-comparison}
  \rho \dd{\entropy}{t} 
  + 
  \divergence 
  \left\{ 
    \entfluxc 
    - 
    \frac{
      \frac{\beta \mu \tilde{\mu}}{4} 
      \Tr
      \left[
        \left( \nabla \lcgII \right) \left( \lcgII - \identity \right)
        +
        \left( \lcgII - \identity \right) \left( \nabla \lcgII \right)
      \right]
    }{
      \temp
    }
  \right\}
  =
  \frac{1}{\temp}
  \Bigg\{
    2 \nu \tensordot{\traceless{\gradsym}}{\traceless{\gradsym}}
    \\
    -
    \frac{\mu}{2}
    \tensordot{
      \left[
        -
        \mu \varphi
        \left(
          \lcgII - \identity
        \right)
        +
        \frac{\beta}{2}
        \left[
          \divergence \left( \tilde{\mu} \nabla \lcgII \right) \lcgII
          +
          \lcgII \divergence \left( \tilde{\mu} \nabla \lcgII \right)
        \right]
      \right]
    }
    {
      \left(
        \identity - \inverse{\lcgII}
      \right)
    }
    +
    \chi
    \frac{\left( \varphi_0 - \varphi \right)^2}{\tau}
    \\
    +
    \frac{
      \vectordot{
        \left[
          \kappa \nabla \temp
          +
          \frac{\beta \mu \tilde{\mu}}{4} 
          \Tr
          \left[
            \left( \nabla \lcgII \right) \left( \lcgII - \identity \right)
            +
            \left( \lcgII - \identity \right) \left( \nabla \lcgII \right)
          \right]
        \right]
      }{
        \nabla \temp
      }
    }{
      \temp
    }
  \Bigg\}.
\end{multline}
Equation \eqref{eq:bmp-entropy-evolution-comparison} can now be compared with the derived entropy evolution equation \eqref{eq:bmp-entropy-evolution-from-fenergy} to yield constitutive relations which guarantee the validity of the second law of thermodynamics.

\subsection{Constitutive relations}
\label{sec:bmp-constitutive-relations}
In particular, equations \eqref{eq:bmp-entropy-evolution-from-fenergy} and \eqref{eq:bmp-entropy-evolution-comparison} will coincide if we set
\begin{subequations}
  \label{eq:bmp-constitutive-relations}
  \begin{align}
  \label{eq:bmp-cstress}
  \traceless{\cstress}
  &=
  2 \nu \traceless{\gradsym}
    +
    \mu \traceless{\left( \lcgII \right)}
    -
    \alpha
    \chi
    \left( \varphi_{0} - \varphi \right)
    \left( \varphi_{\infty} - \varphi \right) \traceless{\gradsym}
    ,
  \\
  \label{eq:bmp-lcgII-evolution-equation}
  \frac{1}{\varphi} \fid{\overline{\lcgII}}
  &=
  -
  \mu
  \left(
    \lcgII - \identity
  \right)
  +
  \frac{\beta}{2 \varphi}
  \left[
    \divergence \left( \tilde{\mu} \nabla \lcgII \right) \lcgII
    +
    \lcgII \divergence \left( \tilde{\mu} \nabla \lcgII \right)
  \right],
  \\
  \label{eq:bmp-varphi}
  \dd{\varphi}{t}
  &=
  \frac{\varphi_0 - \varphi}{\tau} + \alpha \left( \varphi_{\infty} - \varphi \right) \tensordot{\traceless{\gradsym}}{\traceless{\gradsym}},
  \\
  \label{eq:bmp-eflux}
  \efluxc
  &=
    -
    \kappa \nabla \temp
    -
    \frac{\beta \mu \tilde{\mu}}{4} 
    \Tr
    \left[
    \left( \nabla \lcgII \right) \left( \lcgII - \identity \right)
    +
    \left( \lcgII - \identity \right) \left( \nabla \lcgII \right)
    \right]
    ,
    \\
  \label{eq:bmp-entflux}
  \entfluxc
  &=
  -\frac{\kappa \nabla \temp}{\temp}
  ,
  \end{align}
\end{subequations}
where $\alpha$ and $\varphi_{\infty}$ are positive material parameters. Relations \eqref{eq:bmp-constitutive-relations} can then be used for formulating the evolution equations for the mechanical quantities as well as the temperature evolution equation by appealing the standard procedure. Before we formulate the complete set of governing equations, we however need to discuss the positivity of~$\varphi$, which is needed for nonnegativity of the entropy production \emph{ansatz} \eqref{eq:bmp-entropy-production}.

Let us assume that $\varphi_{\infty}> \varphi_0 > 0$, and let us analyse the behaviour of $\varphi$ as it is implied by~\eqref{eq:bmp-varphi}. We see that if~$\varphi < \varphi_0$, then the right-hand side of~\eqref{eq:bmp-varphi} is positive. It means that whenever we have $\varphi < \varphi_0$, then $\varphi$ is an increasing function, and, in particular the value of $\varphi$ can not decrease to zero. On the other hand if $\varphi > \varphi_{\infty}$, then the right-hand side of~\eqref{eq:bmp-varphi} is negative, hence $\varphi$ is a decreasing function. Consequently, if $\varphi \in \left[ \varphi_0, \varphi_{\infty} \right)$, then~\eqref{eq:bmp-varphi} predicts that $\varphi$ will stay in this interval forever. Moreover if $\tensordot{\traceless{\gradsym}}{\traceless{\gradsym}}$ is small then $\varphi$ has a tendency to go to the equilibrium value $\varphi_0$, and, on the other hand, if $\tensordot{\traceless{\gradsym}}{\traceless{\gradsym}}$ is large, then $\varphi$ is driven to $\varphi_{\infty}$. In this sense $\varphi$ indeed works as a variable monitoring  the ``flow-induced microstructural changes'' from the equilibrium value $\varphi_0$ to the nonequilibrium (strong shear flow) value~$\varphi_{\infty}$. A similar analysis can be done in the case $ \varphi_0 > \varphi_{\infty} > 0$. 

We note that~\eqref{eq:bmp-varphi} is the equation that is different from the equation used in the Bautista--Manero--Puig models, since in these models the evolution equation for $\varphi$ usually takes the form 
\begin{equation}
  \label{eq:16}
  \dd{\varphi}{t}
  =
  \frac{\varphi_0 - \varphi}{\tau} + \alpha \left( \varphi_{\infty} - \varphi \right) \tensordot{\traceless{\left(\lcgII\right)}}{\traceless{\gradsym}}.
\end{equation}
In the simple setting outlined above we however cannot guarantee the positivity of the product $\tensordot{\traceless{\left(\lcgII\right)}}{\traceless{\gradsym}}$, and consequently the positivity of $\varphi$, hence we are forced to replace~\eqref{eq:16} by~\eqref{eq:bmp-varphi}.

\subsection{Governing equations}
\label{sec:bmp-governing-equations}
The full system of governing equations for the diffusive Bautista--Manero--Puig type model reads
\begin{subequations}
  \label{eq:bmp-governing-equations}
  \begin{align}
    \label{eq:bmp-incompressibility-condition}
    \divergence \vec{v} 
    &=
    0,
    \\
    \label{eq:bmp-linear-momentum-balance}
    \rho \dd{\vec{v}}{t}
    &=
    \divergence \cstress,
    \\
    \label{eq:bmp-lcg-evolution-equation}
    \frac{1}{\varphi} \fid{\overline{\lcgII}}
    &=
      -
      \mu
      \left(
      \lcgII - \identity
      \right)
      +
      \frac{\beta}{2 \varphi}
      \left[
      \divergence \left( \tilde{\mu} \nabla \lcgII \right) \lcgII
      +
      \lcgII \divergence \left( \tilde{\mu} \nabla \lcgII \right)
      \right],
    \\
    \label{eq:bmp-varphi-evolution-equation}
    \dd{\varphi}{t}
    &=
      \frac{\varphi_0 - \varphi}{\tau} + \alpha \left( \varphi_{\infty} - \varphi \right) \tensordot{\traceless{\gradsym}}{\traceless{\gradsym}},
    \\
    \label{eq:bmp-temperature-evolution-equation}
    \rho \cheatvol^{\mathrm{NSE}}
    \dd{\temp}{t}
    &=
    \divergence \left( \kappa \nabla \temp \right)
      +
      2 \nu \tensordot{\traceless{\gradsym}}{\traceless{\gradsym}}
      +
      \frac{\mu^2 \varphi}{2} 
      \left[ 
      \Tr \lcgII + \Tr \inverse{\lcgII} - 6
      \right]
      +
      \frac{\beta \mu \tilde{\mu}}{2}
      \tensorddot{
      \nabla \lcgII
      }
      {
      \nabla \lcgII
      }
      +
      \chi
      \frac{\left( \varphi_0 - \varphi \right)^2}{\tau}
      ,
  \end{align}
  where the Cauchy stress tensor $\cstress$ is given by the formulae
  \begin{equation}
    \cstress = \mns \identity + \traceless{\cstress},
    \qquad
    \traceless{\cstress} 
    = 
  2 \nu \traceless{\gradsym}
    +
    \mu \traceless{\lcgII}
    -
    \alpha \left( \varphi_{\infty} - \varphi \right) \traceless{\gradsym}
    .
  \end{equation}
\end{subequations}

\section{Implications of the thermodynamical background}
\label{sec:impl-therm-backgr}
Having established the thermodynamical background for various viscoelastic rate-type models with stress diffusion, we can focus on its implications for the analysis of flows of fluids described by these models.

\subsection{Boundary conditions \textendash{} characterisation of thermodynamically isolated or open systems}
\label{sec:boundary-conditions}

First, as a by-product of the thermodynamical analysis, we have obtained explicit expressions for the energy flux and entropy flux. This means that we can say which choice of \emph{boundary conditions} leads to thermodynamically isolated/open systems. Let us consider the fluid occupying the domain $\Omega$. In order to get a thermodynamically isolated system, we have to impose the no-penetration boundary condition on the velocity
\begin{subequations}
  \label{eq:thermodynamically-closed-systems-bc}
  \begin{equation}
    \label{eq:18}
    \left. \vectordot{\vec{v}}{\vec{n}} \right|_{\partial \Omega} = 0,
  \end{equation}
  and the no-mechanical flux boundary condition
  \begin{equation}
    \label{eq:19}
    \left. \vectordot{\cstress \vec{v}}{\vec{n}} \right|_{\partial \Omega} = 0,
  \end{equation}
  where $\vec{n}$ is the unit outward normal to the boundary of $\Omega$. Further we have to enforce the zero flux boundary condition for the non-mechanical contribution to the energy flux $\efluxc$,
  \begin{equation}
    \label{eq:20}
    \left. \vectordot{\efluxc}{\vec{n}} \right|_{\partial \Omega} = 0,
  \end{equation}
\end{subequations}
where $\efluxc$ is given by a model specific formula. (See for example~\eqref{eq:giesekus-eflux} for the non-mechanical energy flux in the diffusive Giesekus model.)

\subsection{Evolution equation for temperature \textendash{}  complete system of governing equations for complex processes}
\label{sec:evol-equat-temp}

Second, for \emph{all the discussed models we have an evolution equation for the temperature that properly takes into account all energy transfer mechanisms in the fluid} described by the given model. This allows one to investigate not only the evolution of mechanical variables but also of the temperature field.  

\subsection{Qualitative behaviour \textendash{}  conserved quantities and stability analysis}
\label{sec:cons-quant-stab}

Third, the outlined analysis also allows one to immediately identify the quantities that are conserved during the motion. Let us for example consider the system of governing equations for the diffusive Johnson--Segalman model in the absence of body forces, that is $\vec{b} = \vec{0}$. The system reads
\begin{subequations}
  \label{eq:26}
  \begin{align}
    \label{eq:27}
    \divergence \vec{v} 
    &=
    0,
    \\
    \label{eq:28}
    \rho \dd{\vec{v}}{t}
    &=
    \divergence \cstress,
    \\
    \label{eq:29}
    \nu_1 \gfid{\overline{\lcgGSII}}
    &=
    -
    \mu
    \left(
      \lcgGSII - \identity
    \right)
    +
    \frac{1}{2}
    \left[
    \divergence \left( \tilde{\mu} \nabla \lcgGSII \right) \lcgGSII
    +
    \lcgGSII \divergence \left( \tilde{\mu} \nabla \lcgGSII \right)
    \right],
    \\
    \label{eq:30}
    \rho \cheatvol^{\mathrm{NSE}}
    \dd{\temp}{t}
    &=
    \divergence \left( \kappa \nabla \temp \right)
    +
    2 \nu \tensordot{\traceless{\gradsym}}{\traceless{\gradsym}}
    +
    \frac{\mu^2}{2 \nu_1} 
    \left[ 
      \Tr \lcgGSII + \Tr \inverse{\lcgGSII} - 6
    \right]
    +
    \frac{\mu \tilde{\mu}(\temp)}{2\nu_1}
    \tensorddot{
      \nabla \lcgGSII
    }
    {
      \nabla \lcgGSII
    },
  \end{align}
  where the Cauchy stress tensor $\cstress$ is given by the formulae
  \begin{equation}
    \label{eq:31}
    \cstress = \mns \identity + \traceless{\cstress},
    \qquad
    \traceless{\cstress} = 2 \nu \gradsym + a \mu \traceless{(\lcgGSII)},
  \end{equation}
\end{subequations}
see~\eqref{eq:js-governing-equations}, and where $\gfid{\overline{\lcgGSII}}$ denotes the Gordon--Schowalter derivative,
\begin{equation}
  \label{eq:23}
  \gfid{\overline{\lcgGSII}}
  =_{\bydefinition}
  \dd{\lcgGSII}{t}
  -
  a
  \left(
    \gradsym
    \lcgGSII
    +
    \lcgGSII
    \gradsym
  \right)
  -
  \left(
    \gradasym
    \lcgGSII
    +
    \lcgGSII
    \transpose{\gradasym}
  \right)
  .
\end{equation}
Given the system of governing equations one can ask the question whether there is a quantity that is conserved in the motion.

We know that in \emph{isolated systems} the quantity that is being conserved is the net total energy. (This means the total energy $\rho \ienergy + \frac{1}{2} \rho \absnorm{\vec{v}}^2$ integrated over the domain $\Omega$.) Having studied the thermodynamical background of the models, we are able to find an explicit formula for the total energy. We know that the Helmholtz free energy is in the case of the diffusive Johnson--Segalman model given by the formula
\begin{equation}
  \label{eq:22}
  \fenergy
  =
  -
  \cheatvol^{\mathrm{NSE}}
  \temp
  \left(
    \ln \frac{\temp}{\tempref} - 1
  \right)
  +
  \frac{\mu}{2\rho}
  \left(
  \Tr \lcgGSII
  -
  3
  -
  \ln \det \lcgGSII
  \right)
  ,
\end{equation}
see~\eqref{eq:js-free-energy} and~\eqref{eq:11}. The internal energy is obtained from the Helmholtz free energy via the formula $\ienergy = \fenergy + \temp \entropy$, where $\entropy = - \pd{\fenergy}{\temp}$, which yields
\begin{equation}
  \label{eq:24}
  \ienergy
  =
  \cheatvol^{\mathrm{NSE}}
  \temp
  +
  \frac{\mu}{2\rho}
  \left(
  \Tr \lcgGSII
  -
  3
  -
  \ln \det \lcgGSII
  \right)
  .
\end{equation}
Consequently, for the diffusive Johnson--Segalman model, we expect that in the isolated system we get
\begin{equation}
  \label{eq:25}
  \dd{}{t}
  \int_{\Omega}
  \left(
    \frac{1}{2}
    \rho
    \absnorm{\vec{v}}^2
    +
    \frac{\mu}{2}
    \left(
      \Tr \lcgGSII
      -
      3
      -
      \ln \det \lcgGSII
    \right)
    +
    \rho
    \cheatvol^{\mathrm{NSE}}
    \temp
  \right)
  \,
  \cvolumee
  =
  0
  .
\end{equation}
Let us see whether this piece of information can be directly derived using the evolution equations only.

We start with a standard manipulation, see for example~\cite{gurtin.me.fried.e.ea:mechanics}. We take the scalar product of the balance of liner momentum~\eqref{eq:28} with the velocity field $\vec{v}$, we integrate the equation over the domain $\Omega$, and we use the integration \emph{by parts} formula, and the no-penetration boundary condition~\eqref{eq:18}, which yields
\begin{equation}
  \label{eq:32}
  \dd{}{t}
  \int_{\Omega}
  \frac{1}{2}
  \rho
  \absnorm{\vec{v}}^2
  \,
  \cvolumee
  =
  -
  \int_{\Omega} \tensordot{\cstress}{\gradsym} \, \cvolumee
  +
  \int_{\Omega} \divergence \left( \cstress \vec{v} \right) \, \cvolumee
  .
\end{equation}
If we take into account the incompressibility condition~\eqref{eq:27} and the structure of the Cauchy stress tensor, see~\eqref{eq:31}, and if we use the Stokes theorem, then we can rewrite~\eqref{eq:32} as
\begin{equation}
  \label{eq:33}
  \dd{}{t}
  \int_{\Omega}
  \frac{1}{2}
  \rho
  \absnorm{\vec{v}}^2
  \,
  \cvolumee
  =
  -
  \int_{\Omega} 2 \nu \tensordot{\gradsym}{\gradsym} \, \cvolumee
  -
  \int_{\Omega} a \mu \tensordot{\lcgGSII}{\gradsym} \, \cvolumee
  +
  \int_{\partial \Omega} \vectordot{\cstress \vec{v}}{\vec{n}} \, \csurfacees
  ,
\end{equation}
where $\vec{n}$ denotes the unit outward normal to the boundary of $\Omega$.

Now take the trace of the evolution equation \eqref{eq:29} for $\lcgGSII$ which, upon using the definition of the Gordon--Schowalter derivative, yields
\begin{equation}
  \label{eq:34}
  \nu_1 \dd{}{t} \Tr \lcgGSII
  -
  2
  a
  \nu_1
  \tensordot{\lcgGSII}{\gradsym}
  =
  -
  \mu
  \left(
    \Tr \lcgGSII - 3
  \right)
  +
  \frac{1}{2}
  \Tr
  \left[
    \divergence \left( \tilde{\mu} \nabla \lcgGSII \right) \lcgGSII
    +
    \lcgGSII \divergence \left( \tilde{\mu} \nabla \lcgGSII \right)
  \right]
  .
\end{equation}
Next we multiply the evolution equation \eqref{eq:29} for $\lcgGSII$ by $\inverse{\lcgGSII}$ and then we take the trace. This manipulation yields
\begin{equation}
  \label{eq:35}
  \nu_1 \dd{}{t} \ln \det \lcgGSII
  =
  -
  \mu
  \left(
    3 -  \Tr \inverse{\lcgGSII}
  \right)
  +
  \frac{1}{2}
  \Tr
  \left(
    \left[
      \divergence \left( \tilde{\mu} \nabla \lcgGSII \right) \lcgGSII
      +
      \lcgGSII \divergence \left( \tilde{\mu} \nabla \lcgGSII \right)
    \right]
    \inverse{\lcgGSII}
  \right)
  ,
\end{equation}
where we have used the incompressibility constraint $\divergence \vec{v} = \Tr \gradsym = 0$, and the well known formula $\dd{}{t} \ln \det \generictensor = \Tr \left( \dd{\generictensor}{t} \inverse{\generictensor} \right)$. Subtracting~\eqref{eq:35} from~\eqref{eq:34} and integrating over the domain $\Omega$ finally leads to the equation
\begin{multline}
  \label{eq:36}
  \int_{\Omega}
  \left(
    \nu_1
    \dd{}{t}
    \left(
      \Tr \lcgGSII
      -
      \ln \det \lcgGSII
    \right)
    -
    2
    a
    \nu_1
    \tensordot{\lcgGSII}{\gradsym}
  \right)
  \,
  \cvolumee
  =
  -
  \int_{\Omega}
  \mu
  \left(
    \Tr \lcgGSII + \Tr \inverse{\lcgGSII} - 6
  \right)
  \,
  \cvolumee
  \\
  +
  \int_{\Omega}
  \frac{1}{2}
  \Tr
  \left(
    \left[
      \divergence \left( \tilde{\mu} \nabla \lcgGSII \right) \lcgGSII
      +
      \lcgGSII \divergence \left( \tilde{\mu} \nabla \lcgGSII \right)
    \right]
    \left(
      \identity
      -
      \inverse{\lcgGSII}
    \right)
  \right)
  \,
  \cvolumee
  .
\end{multline}

Having derived~\eqref{eq:36}, we can multiply~\eqref{eq:36} by $\frac{\mu}{2\nu_1}$ and add it to~\eqref{eq:32}, which yields
\begin{multline}
  \label{eq:37}
  \dd{}{t}
  \int_{\Omega}
  \left(
    \frac{1}{2}
    \rho \absnorm{\vec{v}}^2
    +
    \frac{\mu}{2}
    \left(
      \Tr \lcgGSII
      -
      3
      -
      \ln \det \lcgGSII
    \right)
  \right)
  \,
  \cvolumee
  =
  -
  \int_{\Omega}
  2
  \nu
  \tensordot{\gradsym}{\gradsym}
  \,
  \cvolumee
  -
  \int_{\Omega}
  \frac{\mu^2}{2 \nu_1}
  \left(
    \Tr \lcgGSII + \Tr \inverse{\lcgGSII} - 6
  \right)
  \,
  \cvolumee
  \\
  +
  \frac{\mu}{2 \nu_1}
  \int_{\Omega}
  \frac{1}{2}
  \Tr
  \left(
    \left[
      \divergence \left( \tilde{\mu} \nabla \lcgGSII \right) \lcgGSII
      +
      \lcgGSII \divergence \left( \tilde{\mu} \nabla \lcgGSII \right)
    \right]
    \left(
      \identity
      -
      \inverse{\lcgGSII}
    \right)
  \right)
  \,
  \cvolumee
  +
  \int_{\partial \Omega} \vectordot{\cstress \vec{v}}{\vec{n}} \, \csurfacees
  .
\end{multline}
(On the left hand side we have added constant $3$ to the term under the time derivative, which is an operation that does not change the resulting equation. The constant is added for the matter of convenience, it basically fixes the energy of the equilibrium state $\lcgGSII = \identity$ to zero.) Using \eqref{eq:tensorddot-identity} we rewrite the terms on the right hand side, and we get
\begin{multline}
  \label{eq:38}
  \dd{}{t}
  \int_{\Omega}
  \left(
    \frac{1}{2}
    \rho \absnorm{\vec{v}}^2
    +
    \frac{\mu}{2}
    \left(
      \Tr \lcgGSII
      -
      3
      -
      \ln \det \lcgGSII
    \right)
  \right)
  \,
  \cvolumee
  \\
  =
  -
  \int_{\Omega}
  2
  \nu
  \tensordot{\gradsym}{\gradsym}
  \,
  \cvolumee
  -
  \int_{\Omega}
  \frac{\mu^2}{2 \nu_1}
  \left(
    \Tr \lcgGSII + \Tr \inverse{\lcgGSII} - 6
  \right)
  \,
  \cvolumee
  -
  \int_{\Omega}
  \frac{\mu \tilde{\mu}}{2\nu_1}
  \tensorddot{
    \nabla \lcgGSII
  }
  {
    \nabla \lcgGSII
  }
  \,
  \cvolumee
  \\
  +
  \int_{\Omega}
  \divergence
  \left(
    \frac{\mu \tilde{\mu}}{4 \nu_1}
    \Tr
    \left[
      \left( \nabla \lcgGSII \right) \left( \lcgGSII - \identity \right)
      +
      \left( \lcgGSII - \identity \right) \left( \nabla \lcgGSII \right)
    \right]
  \right)
  \,
  \cvolumee
  +
  \int_{\partial \Omega}
  \vectordot{\cstress \vec{v}}{\vec{n}}
  \,
  \csurfacees
  .
\end{multline}
The next to the last integral can be converted to a surface integral using the Stokes theorem. Moreover, we see that the term under the divergence can be rewritten as $- \efluxc - \kappa \nabla \temp$, see \eqref{eq:js-eflux}, which yields
\begin{multline}
  \label{eq:40}
  \dd{}{t}
  \int_{\Omega}
  \left(
    \frac{1}{2}
    \rho \absnorm{\vec{v}}^2
    +
    \frac{\mu}{2}
    \left(
      \Tr \lcgGSII
      -
      3
      -
      \ln \det \lcgGSII
    \right)
  \right)
  \,
  \cvolumee
  \\
  =
  -
  \int_{\Omega}
  2
  \nu
  \tensordot{\gradsym}{\gradsym}
  \,
  \cvolumee
  -
  \int_{\Omega}
  \frac{\mu^2}{2 \nu_1}
  \left(
    \Tr \lcgGSII + \Tr \inverse{\lcgGSII} - 6
  \right)
  \,
  \cvolumee
  -
  \int_{\Omega}
  \frac{\mu \tilde{\mu}}{2\nu_1}
  \tensorddot{
    \nabla \lcgGSII
  }
  {
    \nabla \lcgGSII
  }
  \,
  \cvolumee
  \\
  -
  \int_{\partial \Omega}
  \vectordot{\efluxc}{\vec{n}}
  \,
  \csurfacees
  -
  \int_{\partial \Omega}
  \vectordot{\kappa \nabla \temp}{\vec{n}}
  \,
  \csurfacees
  +
  \int_{\partial \Omega}
  \vectordot{\cstress \vec{v}}{\vec{n}}
  \,
  \csurfacees
  .
\end{multline}
If we are dealing with an isolated system (no work is done on the system, no heat flux into the system, no energy flux into the system), then all the surface integrals vanish a we see that the quantity
\begin{equation}
  \label{eq:41}
  \int_{\Omega}
  \left(
    \frac{1}{2}
    \rho \absnorm{\vec{v}}^2
    +
    \frac{\mu}{2}
    \left(
      \Tr \lcgGSII
      -
      3
      -
      \ln \det \lcgGSII
    \right)
  \right)
  \,
  \cvolumee
\end{equation}
decays in time. (All the remaining terms on the right hand side are negative.) This is not surprising provided the we know the thermodynamical background of the model.

In fact we could have seen it \emph{a priori} without any manipulation of the governing equations. The fact that the quantity given by~\eqref{eq:41} decays in time is a consequence of the design of the model. In a thermodynamically isolated system the net \emph{mechanical energy}, which is precisely the quantity given by the integral~\eqref{eq:41}, must degrade to the thermal energy. Note that this simple consequence of the thermodynamical background of the model was rather difficult to obtain directly from the governing equations.

Let us further investigate~\eqref{eq:40}. Integrating the evolution equation for the temperature, see~\eqref{eq:30}, and adding the result to~\eqref{eq:40} we get
\begin{equation}
  \label{eq:42}
  \dd{}{t}
  \int_{\Omega}
  \left(
    \frac{1}{2}
    \rho \absnorm{\vec{v}}^2
    +
    \frac{\mu}{2}
    \left(
      \Tr \lcgGSII
      -
      3
      -
      \ln \det \lcgGSII
    \right)
    +
    \rho \cheatvol^{\mathrm{NSE}} \temp
  \right)
  \,
  \cvolumee
  \\
  =
  \int_{\partial \Omega}
  \vectordot{\cstress \vec{v}}{\vec{n}}
  \,
  \csurfacees
  -
  \int_{\partial \Omega}
  \vectordot{\efluxc}{\vec{n}}
  \,
  \csurfacees
  .
\end{equation}
where we have again used the Stokes theorem. The equality tells us that the quantity on the left-hand side is conserved in thermodynamically isolated systems, that is in systems with boundary conditions specified via~\eqref{eq:thermodynamically-closed-systems-bc}. If~\eqref{eq:thermodynamically-closed-systems-bc} holds, then~\eqref{eq:42} coincides with~\eqref{eq:25} which automatically follows from the thermodynamical background of the model. Manipulations of this type are essential in the mathematical analysis of the corresponding models, see for example~\cite{bulvcek.m.malek.j.ea:pde-analysis,bul-cek.m.feireisl.e.ea:on}.

Besides the automatic identification of conserved quantities in thermodynamically isolated systems, the knowledge of thermodynamical background of the models can be also exploited in the stability analysis. For example, if we consider a thermodynamically isolated system such as a fluid in a thermodynamically isolated vessel, then we expect that the fluid will eventually reach a homogeneous steady state regardless of the initial state of the fluid. (Here homogeneous steady state means zero velocity field, $\lcgGSII = \identity$ and spatially homogeneous temperature field.) This natural tendency would be rather difficult to see directly from the governing equations.

However, since we know the thermodynamical background of the models, we know that ``entropy rises'' and ``energy is conserved''. Consequently, we can design a functional that can serve as a Lyapunov-like functional that charaterises the approach to the homogeneous steady state, see \cite{coleman.bd:on} and \cite{gurtin.me:thermodynamics*1,gurtin.me:thermodynamics} for early studies in this direction. The  Lyapunov-like functional is made up of the net entropy $\netentropy$ and the net total energy $\nettenergy$
\begin{subequations}
  \label{eq:45}
  \begin{align}
    \label{eq:44}
    \netentropy
    &=_{\bydefinition}
      \int_{\Omega}
      \rho
      \entropy
      \,
      \cvolumee
      ,
    \\
    \label{eq:46}
    \nettenergy
    &=_{\bydefinition}
      \int_{\Omega}
      \left(
      \rho
      \ienergy
      +
      \frac{1}{2} \rho \absnorm{\vec{v}}^2
      \right)
      \,
      \cvolumee
      ,
  \end{align}
\end{subequations}
and it is given by the formula
\begin{equation}
  \label{eq:39}
  \mathcal{V}_{\temp_{\reference}, \mathrm{eq}} 
  =_{\bydefinition} 
  -
  \left[
    \netentropy - \frac{1}{\temp_{\reference}} \left( \nettenergy - \widehat{\nettenergy} \right)
  \right]
  ,
\end{equation}
where $\temp_{\reference}$ is the temperature at the steady state and $\widehat{\nettenergy}$ is the energy at the steady state (a known constant), see~\cite{bul-cek.m.malek.j.ea:thermodynamics} for details. Fortunately, since we have completely explored the thermodynamical background of the models, we know the explicit formulae for the internal energy $\ienergy$ and the specific entropy $\entropy$ for the given models, hence we can explicitly construct the functional.

Taking the time derivative of the functional $\mathcal{V}_{\temp_{\reference}, \mathrm{eq}}$ yields
\begin{equation}
  \label{eq:43}
  \dd{}{t}
  \mathcal{V}_{\temp_{\reference}, \mathrm{eq}}
  = 
  -
  \dd{}{t}
  \left[ 
    \netentropy - \frac{1}{\temp_{\reference}} \left( \nettenergy - \widehat{\nettenergy} \right) 
  \right]
  \\
  =
  -
  \dd{\netentropy}{t}
  =
  -
  \int_{\Omega} \entprodc
  \leq
  0
  ,
\end{equation}
where we have exploited the fact that the net total energy is conserved in a thermodynamically isolated system, $\dd{\nettenergy}{t} = 0$, and that the net entropy grows in a thermodynamically isolated system. (The growth of the net entropy follows by integration of the evolution equation~\eqref{eq:entropy-evolution-general} the domain $\Omega$, the positivity of the entropy production term $\entprodc$ and the no-flux boundary condition for the entropy flux.) This simple manipulation shows that the functional indeed decreases in time hence it is truly a Lyapunov-like functional.

The reader interested in the application of this procedure in the context of viscoelastic rate-type fluids is kindly referred to~\cite{malek.j.prusa.v.ea:thermodynamics}. In this work the authors have analysed the stability of the equilibrium steady state of a fluid described by a diffusive variant of Oldroyd-B model. The successful identification of the energy storage and entropy producing mechanisms for other diffusive viscoelastic-type models, see above, opens up the possibility to redo the stability analysis by~\cite{malek.j.prusa.v.ea:thermodynamics} for these models as well.

Thermodynamical background can be however exploited in more complex settings than the stability analysis of the rest state. In fact, thermodynamical background is of use even in the \emph{stability analysis of steady states in thermodynamically open systems}, see~\cite{bul-cek.m.malek.j.ea:thermodynamics}. \cite{bul-cek.m.malek.j.ea:thermodynamics} have conjectured, that the functional
\begin{equation}
  \label{eq:47}
  \mathcal{V}_{\mathrm{neq}}
  \left(
    \left.
      \widetilde{\vec{W}}
    \right\|
    \widehat{\vec{W}}
  \right)
  =_{\bydefinition}
  -
  \left\{
    \entropyrellp_{\widehat{\temp}}(\left. \widetilde{\vec{W}} \right\| \widehat{\vec{W}})
    -
    \energyrellp  (\left. \widetilde{\vec{W}} \right\| \widehat{\vec{W}})
  \right\}
\end{equation}
where
\begin{subequations}
  \label{eq:48}
  \begin{align}
    \label{eq:49}
    \entropyrellp_{\widehat{\temp}}(\left. \widetilde{\vec{W}} \right\| \widehat{\vec{W}})
    &=
    _{\bydefinition}
    \netentropy_{\widehat{\temp}} 
    \left(
      \widehat{\vec{W}}
      +
      \widetilde{\vec{W}}
    \right)
    -
    \netentropy_{\widehat{\temp}} 
    \left(
      \widehat{\vec{W}}
    \right)
    -
    \left.
      \Diff[\vec{W}] \netentropy_{\widehat{\temp}}
      \left(
        \vec{W}
      \right)
    \right|_{\vec{W}  = \widehat{\vec{W}}}
    \left[
      \widetilde{\vec{W}}
      \right]
      ,
    \\
    \label{eq:50}
    \energyrellp  (\left. \widetilde{\vec{W}} \right\| \widehat{\vec{W}})
    &=
    _{\bydefinition}
    \nettenergy 
    \left(
      \widehat{\vec{W}}
      +
      \widetilde{\vec{W}}
    \right)
    -
    \nettenergy 
    \left(
      \widehat{\vec{W}}
    \right)
    -
    \left.
      \Diff[\vec{W}] \nettenergy
      \left(
        \vec{W}
      \right)
    \right|_{\vec{W}  = \widehat{\vec{W}}}
    \left[
      \widetilde{\vec{W}}
      \right]
      ,
    \\
    \label{eq:51}
    \netentropy_{\widehat{\temp}} \left(\vec{W}\right)
    &=
    _{\bydefinition}
    \int_{\Omega}
    \rho
    \widehat{\temp}
    \entropy(\vec{W})
    \,
      \cvolumee
      ,
    \\
    \label{eq:52}
    \nettenergy\left(\vec{W}\right)
    &=
    _{\bydefinition}
    \int_{\Omega}
    \rho
    \ienergy(\vec{W})
    \,
      \cvolumee
      .
\end{align}
\end{subequations}
could serve as a Lyapunov-like functional for the analysis of the steady state in a thermodynamically open system. Here $\widehat{\vec{W}} = \left[ \widehat{\vec{v}}, \widehat{\lcgGSII}, \widehat{\temp} \right]$ denotes the vector of state variables in the steady state, and $\widetilde{\vec{W}} = \left[ \widetilde{\vec{v}}, \widetilde{\lcgGSII}, \widetilde{\temp} \right]$ denotes the vector of the perturbations of the state variables with respect to the steady state, and
$
\left.
  \Diff[\vec{W}] \netentropy_{\widehat{\temp}}
  \left(
    \vec{W}
  \right)
\right|_{\vec{W}  = \widehat{\vec{W}}}
\left[
  \widetilde{\vec{W}}
\right]
$
and
$
\left.
      \Diff[\vec{W}] \nettenergy
      \left(
        \vec{W}
      \right)
    \right|_{\vec{W}  = \widehat{\vec{W}}}
    \left[
      \widetilde{\vec{W}}
      \right]
$
denote the G\^ateaux derivative\footnote{%
\label{fn:1}
Let us recall that the G\^ateaux derivative $\Diff \mathcal{M} (\vec{x})[\vec{y}]$ of a functional $\mathcal{M}$ at point $\vec{x}$ in the direction $\vec{y}$ is defined as
$
    \Diff \mathcal{M} (\vec{x})[\vec{y}]
    =_{\bydefinition}
    \lim_{s \to 0}
    \frac{
      \mathcal{M} (\vec{x} + s \vec{y}) - \mathcal{M} (\vec{x})
    }
    {
      s
    }
$ 
which is tantamount to 
$
\Diff \mathcal{M} (\vec{x})[\vec{y}]
=_{\bydefinition}
\left.
  \dd{}{s}
  \mathcal{M} (\vec{x} + s \vec{y})
\right|_{s=0}
$. If it is necessary to emphasize the variable against which we differentiate, we also write $\Diff[\vec{x}] \mathcal{M} (\vec{x})[\vec{y}]$ instead of $\Diff \mathcal{M} (\vec{x})[\vec{y}]$.
} 
of the given functional with respect to $\vec{W}$ at point $\widehat{\vec{W}}$ in the direction
$
\widetilde{\vec{W}}
$%
,
see~\cite{bul-cek.m.malek.j.ea:thermodynamics} for details.

As in the previous case \emph{the critical piece of information necessary to build the Lyapunov-like functionals is the knowledge of the Helmholtz free energy}, which allows one to obtain the explicit formulae for the entropy and the internal energy, and hence the explicit formulae for the functionals. A reader interested in the application of this procedure in the context of viscoelastic rate-type fluids is kindly referred to~\cite{dostalk.m.prusa.v.ea:finite}, who have investigated the stability of a steady flow of a fluid described by the Giesekus model. The successful identification of the energy storage and entropy producing mechanisms for the diffusive viscoelastic-type models, see above, opens up the possibility to redo the stability analysis by~\cite{dostalk.m.prusa.v.ea:finite} for these models as well.

\section{Conclusion}
\label{sec:conclusion}

We have presented a purely phenomenological approach to the viscoelastic rate-type models with a stress diffusion term. In particular we have identified \emph{energy storage mechanisms} and \emph{entropy production mechanisms} that lead to the diffusive variants of the Giesekus, FENE-P, Johnson--Segalman, Phan-Thien--Tanner and Bautista--Manero--Puig models. In all these cases the stress diffusion term has been interpreted as a consequence of a nonstandard entropy production mechanism. Namely we have assumed that the entropy production contains a quadratic term in the gradient of the ``extra stress tensor'', which in turn have lead to the presence of the stress diffusion term in the corresponding governing equation.

All the models have been based on the knowledge of the triad \emph{energy storage mechanisms}/\emph{entropy production mechanisms}/\emph{underlying kinematics}. The governing equations---\emph{including the evolution equation for the temperature field}---have been interpreted as a simple consequence of the specifications of the individual components of the triad. The nature of the outlined procedure allows one to freely choose individual components of the triad, and alternatively derive more complicated models. For example, the simple quadratic term in the gradient of the ``extra stress tensor'' which appears in the entropy production, can be replaced, if necessary, by a more complicated formula. The same holds for the thermal part of the Helmholtz free energy/entropy production that has been for simplicity chosen in the form that implies constant specific heat capacity and the standard Fourier's law for the heat flux.

Finally, we have discussed the potential use of the provided thermodynamical background in the study of stability of flows of fluids described by the proposed models.






\begin{theacknowledgments}
  Mark Dostal\'{\i}k and V\'{\i}t Pr\r{u}\v{s}a acknowledge the support of the Czech Science Foundation project 18-12719S. Mark Dostal\'{\i}k has also been supported by Charles University Research programme No. UNCE/SCI/023.
\end{theacknowledgments}

\appendix

\section{Corrections}
\label{sec:corrections}

\subsection{Helmholtz free energy for the FENE-P model}
\label{sec:helmh-free-energy}
The Helmholtz free energy \emph{ansatz} for the FENE-P model should read
\begin{equation}
  \label{eq:55}
    \fenergy
    =_{\bydefinition}
    \fenergy_0(\temp)
    + 
    \frac{\mu}{2\rho}
    \left[
    -
    b \ln \left( 1 - \frac{1}{b} \Tr \lcgII \right) 
    +
    b \ln \left( 1 - \frac{3}{b \left( 1 + \frac{3}{b} \right)} \right)
    +
    3
    \ln
    \left(
      \frac{1}{1 + \frac{3}{b}}
    \right)
    -
    \ln \det \lcgII
    \right]
  \end{equation}
  instead of~\eqref{eq:fene-p-free-energy} which reads
  \begin{equation}
    \label{eq:54}
    \fenergy
    =_{\bydefinition}
    \fenergy_0(\temp)
    + 
    \frac{\mu}{2\rho}
    \left[
    -
    b \ln \left( 1 - \frac{1}{b} \Tr \lcgII \right) 
    -
    3
    -
    \ln \det \lcgII
    \right]
  \end{equation}
  The change of the additive constant \emph{has no impact on the constitutive relations and the governing equations}, since these depend on the derivatives of the Helmholtz free energy.

  On the other hand, it is convenient to choose the Helmholtz free energy such that it vanishes at the equilibrium steady state. If the velocity field $\vec{v}$ vanishes, then~\eqref{eq:fene-p-lcg-evolution-equation} implies that the equilibrium steady state value of $\lcgII$ is
  \begin{equation}
    \label{eq:56}
    {\lcgII}_{\mathrm{eq}}
    =
    \frac{
      \identity
    }
    {
      1+ \frac{3}{b}
    }
    .
  \end{equation}
  Consequently if we want the (mechanical) part of the Helmholtz free energy to vanish for ${\lcgII}_{\mathrm{eq}}$, we have to replace~\eqref{eq:54} with~\eqref{eq:55}. The fact that the mechanical part of~\eqref{eq:55} vanishes for ${\lcgII}_{\mathrm{eq}}$ given by~\eqref{eq:56} is straightforward to show by direct substitution. Moreover, if we take the formal limit $b \to + \infty$, we see that the Helmholtz free energy \emph{ansatz} for the FENE-P model reduces to the standard ansatz for the Oldroyd-B model.

  \subsection{Entropy production for the Giesekus model}
  \label{sec:entorpy-prod-gies}
There is a misprint in the entropy production formula~\eqref{eq:giesekus-entropy-production} for the Giesekus model. The wrong formula reads
\begin{equation}
  \label{eq:58}
  \entprodc
  =
  \frac{1}{\temp}
  \left\{
  2 \nu \tensordot{\traceless{\gradsym}}{\traceless{\gradsym}}
  +
  \frac{\mu^2}{2 \nu_1} \Tr 
  \left[ 
    \lambda \lcgII^2 + (1 - 3 \lambda) \lcgII + (1 + \lambda) \inverse{\lcgII} + (3 \lambda - 2) \identity
  \right]
  +
  \frac{\mu \tilde{\mu}(\temp)}{2\nu_1}
  \tensorddot{
    \nabla \lcgII
  }
  {
    \nabla \lcgII
  }
  +
  \kappa
  \vectordot{\nabla \temp}{\frac{\nabla \temp}{\temp}}
\right\}
,
\end{equation}
while the \emph{correct} formula is
\begin{equation}
  \label{eq:59}
  \entprodc
  =
  \frac{1}{\temp}
  \left\{
  2 \nu \tensordot{\traceless{\gradsym}}{\traceless{\gradsym}}
  +
  \frac{\mu^2}{2 \nu_1} \Tr 
  \left[ 
    \lambda \lcgII^2 + (1 - 3 \lambda) \lcgII + (1 - \lambda) \inverse{\lcgII} + (3 \lambda - 2) \identity
  \right]
  +
  \frac{\mu \tilde{\mu}(\temp)}{2\nu_1}
  \tensorddot{
    \nabla \lcgII
  }
  {
    \nabla \lcgII
  }
  +
  \kappa
  \vectordot{\nabla \temp}{\frac{\nabla \temp}{\temp}}
\right\}
.
\end{equation}
The misprint is in the term $(1 + \lambda) \inverse{\lcgII}$ which should read $(1 - \lambda) \inverse{\lcgII}$. The same misprint appears in \eqref{eq:53} and \eqref{eq:giesekus-temperature-evolution-equation} and in Table~\ref{tab:models}.

\begin{thebibliography}{64}
\expandafter\ifx\csname natexlab\endcsname\relax\def\natexlab#1{#1}\fi
\providecommand{\enquote}[1]{``#1''}
\expandafter\ifx\csname url\endcsname\relax
  \def\url#1{\texttt{#1}}\fi
\expandafter\ifx\csname urlprefix\endcsname\relax\def\urlprefix{URL }\fi
\providecommand{\eprint}[2][]{\url{#2}}

\bibitem[Cates and Fielding(2006)]{cates.me.fielding.sm:rheology}
M.~E. Cates, and S.~M. Fielding, \emph{Adv. Phys.} \textbf{55}, 799--879
  (2006).

\bibitem[Subbotin et~al.(2011)]{subbotin.av.malkin.ay.ea:self-organization}
A.~V. Subbotin, A.~Y. Malkin, and V.~G. Kulichikhin, \emph{Adv. Colloid
  Interface Sci.} \textbf{162}, 29--38 (2011).

\bibitem[Fardin et~al.(2015)]{fardin.m.radulescu.o.ea:stress}
M.-A. Fardin, O.~Radulescu, A.~Morozov, O.~Cardoso, J.~Browaeys, and
  S.~Lerouge, \emph{J. Rheol.} \textbf{59}, 1335--1362 (2015).

\bibitem[Olmsted et~al.(2000)]{olmsted.pd.radulescu.o.ea:johnson-segalman}
P.~D. Olmsted, O.~Radulescu, and C.-Y.~D. Lu, \emph{J. Rheol.} \textbf{44},
  257--275 (2000).

\bibitem[Johnson and Segalman(1977)]{johnson.mw.segalman.d:model}
M.~W. Johnson, and D.~Segalman, \emph{J. Non-Newton. Fluid Mech.} \textbf{2},
  255--270 (1977).

\bibitem[Helgeson
  et~al.(2009{\natexlab{a}})]{helgeson.me.vasquez.pa.ea:rheology}
M.~E. Helgeson, P.~A. Vasquez, E.~W. Kaler, and N.~J. Wagner, \emph{J. Rheol.}
  \textbf{53}, 727--756 (2009{\natexlab{a}}).

\bibitem[Helgeson
  et~al.(2009{\natexlab{b}})]{helgeson.me.reichert.md.ea:relating}
M.~E. Helgeson, M.~D. Reichert, Y.~T. Hu, and N.~J. Wagner, \emph{Soft Matter}
  \textbf{5}, 3858--3869 (2009{\natexlab{b}}).

\bibitem[Cheng et~al.(2017)]{cheng.p.burroughs.mc.ea:distinguishing}
P.~Cheng, M.~C. Burroughs, L.~G. Leal, and M.~E. Helgeson, \emph{Rheol. Acta}
  \textbf{56}, 1007--1032 (2017).

\bibitem[Giesekus(1982)]{giesekus.h:simple}
H.~Giesekus, \emph{J. Non-Newton. Fluid Mech.} \textbf{11}, 69--109 (1982).

\bibitem[Adams et~al.(2011)]{adams.jm.fielding.sm.ea:transient}
J.~M. Adams, S.~M. Fielding, and P.~D. Olmsted, \emph{J. Rheol.} \textbf{55},
  1007--1032 (2011).

\bibitem[Chung et~al.(2011)]{chung.c.uneyama.t.ea:numerical}
C.~Chung, T.~Uneyama, Y.~Masubuchi, and H.~Watanabe, \emph{Rheol. Acta}
  \textbf{50}, 753--766 (2011).

\bibitem[Carter et~al.(2016)]{carter.ka.girkin.jm.ea:shear}
K.~A. Carter, J.~M. Girkin, and S.~M. Fielding, \emph{J. Rheol.} \textbf{60},
  883--904 (2016).

\bibitem[Likhtman and Graham(2003)]{likhtman.ae.graham.rs:simple}
A.~E. Likhtman, and R.~S. Graham, \emph{J. Non-Newton. Fluid Mech.}
  \textbf{114}, 1--12 (2003).

\bibitem[El-Kareh and Leal(1989)]{el-kareh.aw.leal.lg:existence}
A.~W. El-Kareh, and L.~G. Leal, \emph{J. Non-Newton. Fluid Mech.} \textbf{33},
  257--287 (1989).

\bibitem[Divoux et~al.(2016)]{divoux.t.fardin.ma.ea:shear}
T.~Divoux, M.~A. Fardin, S.~Manneville, and S.~Lerouge, \emph{Annu. Rev. Fluid
  Mech.} \textbf{48}, 81--103 (2016).

\bibitem[Thomases(2011)]{thomases.b:analysis}
B.~Thomases, \emph{J. Non-Newton. Fluid Mech.} \textbf{166}, 1221--1228 (2011).

\bibitem[Barrett and Boyaval(2011)]{barrett.jw.boyaval.s:existence}
J.~W. Barrett, and S.~Boyaval, \emph{Math. Models Methods Appl. Sci.}
  \textbf{21}, 1783--1837 (2011).

\bibitem[Chupin and Martin(2015)]{chupin.l.martin.s:stationary}
L.~Chupin, and S.~Martin, \emph{J. Non-Newton. Fluid Mech.} \textbf{218},
  27--39 (2015).

\bibitem[Chupin et~al.(2018)]{chupin.l.ichim.a.ea:stationary}
L.~Chupin, A.~Ichim, and S.~Martin, \emph{Z. Angew. Math. Mech.} \textbf{98},
  147--172 (2018).

\bibitem[Luk\'a\v{c}ov\'a-Medvid'ov\'a
  et~al.(2016)]{lukacov-medvidova.m.notsu.h.ea:energy}
M.~Luk\'a\v{c}ov\'a-Medvid'ov\'a, H.~Notsu, and B.~She, \emph{Int. J. Numer.
  Meth. Fluids} \textbf{81}, 523--557 (2016).

\bibitem[Mohammadigoushki and Muller(2016)]{mohammadigoushki.h.muller.sj:flow}
H.~Mohammadigoushki, and S.~J. Muller, \emph{Soft Matter} \textbf{12},
  1051--1061 (2016).

\bibitem[Wapperom and Hulsen(1998)]{wapperom.p.hulsen.ma:thermodynamics}
P.~Wapperom, and M.~A. Hulsen, \emph{J. Rheol.} \textbf{42}, 999--1019 (1998).

\bibitem[Dressler et~al.(1999)]{dressler.m.edwards.bj.ea:macroscopic}
M.~Dressler, B.~J. Edwards, and H.~C. {\"O}ttinger, \emph{Rheol. Acta}
  \textbf{38}, 117--136 (1999).

\bibitem[Rajagopal and
  Srinivasa(2000)]{rajagopal.kr.srinivasa.ar:thermodynamic}
K.~R. Rajagopal, and A.~R. Srinivasa, \emph{J. Non-Newton. Fluid Mech.}
  \textbf{88}, 207--227 (2000).

\bibitem[Ellero et~al.(2003)]{ellero.m.espanol.p.ea:thermodynamically}
M.~Ellero, P.~Espa\~nol, and E.~G. Flekk\o{}y, \emph{Phys. Rev. E} \textbf{68},
  041504 (2003).

\bibitem[Pavelka et~al.(2018)]{pavelka.m.klika.v.ea:multiscale}
M.~Pavelka, V.~Klika, and M.~Grmela, \emph{Multiscale Thermo-Dynamics}, de
  Gruyter, Berlin, 2018.

\bibitem[Manero et~al.(2007)]{manero.o.perez-lopez.jh.ea:thermodynamic}
O.~Manero, J.~H. P\'erez-L\'opez, J.~I. Escalante, J.~E. Puig, and F.~Bautista,
  \emph{J Non-Newton. Fluid Mech.} \textbf{146}, 22--29 (2007).

\bibitem[Vasquez et~al.(2007)]{vasquez.pa.mckinley.gh.ea:network}
P.~A. Vasquez, G.~H. McKinley, and L.~P. Cook, \emph{J. Non-Newton. Fluid
  Mech.} \textbf{144}, 122--139 (2007).

\bibitem[Grmela et~al.(2010)]{grmela.m.chinesta.f.ea:mesoscopic}
M.~Grmela, F.~Chinesta, and A.~Ammar, \emph{Rheol. Acta} \textbf{49}, 495--506
  (2010).

\bibitem[Germann et~al.(2013)]{germann.n.cook.lp.ea:nonequilibrium}
N.~Germann, L.~P. Cook, and A.~N. Beris, \emph{J. Non-Newtonian Fluid Mech.}
  \textbf{196}, 51--57 (2013).

\bibitem[M\'{a}lek et~al.(2018)]{malek.j.prusa.v.ea:thermodynamics}
J.~M\'{a}lek, V.~Pr\r{u}\v{s}a, T.~Sk\v{r}ivan, and E.~S\"{u}li, \emph{Phys.
  Fluids} \textbf{30}, 023101 (2018).

\bibitem[M\'alek et~al.(2015{\natexlab{a}})]{malek.j.rajagopal.kr.ea:on}
J.~M\'alek, K.~R. Rajagopal, and K.~T\r{u}ma, \emph{Int. J. Non-Linear Mech.}
  \textbf{76}, 42--47 (2015{\natexlab{a}}).

\bibitem[Hron et~al.(2017)]{hron.j.milos.v.ea:on}
J.~Hron, V.~Milo\v{s}, V.~Pr\r{u}\v{s}a, O.~Sou\v{c}ek, and K.~T\r{u}ma,
  \emph{Int. J. Non-Linear Mech.} \textbf{95}, 193--208 (2017).

\bibitem[M\'alek and Pr\r{u}\v{s}a(2017)]{malek.j.prusa.v:derivation}
J.~M\'alek, and V.~Pr\r{u}\v{s}a, \enquote{Derivation of equations for
  continuum mechanics and thermodynamics of fluids,} in \emph{Handbook of
  Mathematical Analysis in Mechanics of Viscous Fluids}, edited by Y.~Giga, and
  A.~Novotn\'y, Springer, 2017, pp. 1--70.

\bibitem[Rajagopal and Srinivasa(2004)]{rajagopal.kr.srinivasa.ar:on*7}
K.~R. Rajagopal, and A.~R. Srinivasa, \emph{Proc. R. Soc. Lond., Ser. A, Math.
  Phys. Eng. Sci.} \textbf{460}, 631--651 (2004).

\bibitem[Hu and Leli{\`e}vre(2007)]{hu.d.lelievre.t:new}
D.~Hu, and T.~Leli{\`e}vre, \emph{Commun. Math. Sci.} \textbf{5}, 909--916
  (2007).

\bibitem[Barrett and Boyaval(2017)]{barrett.j.boyaval.s:finite}
J.~W. Barrett, and S.~Boyaval, \emph{IMA J. Numer. Anal.} p. drx061 (2017).

\bibitem[Gordon and Schowalter(1972)]{gordon.rj.schowalter.wr:anisotropic}
R.~J. Gordon, and W.~R. Schowalter, \emph{Trans. Soc. Rheol.} \textbf{16},
  79--97 (1972).

\bibitem[Wineman and Rajagopal(2000)]{wineman.as.rajagopal.kr:mechanical}
A.~S. Wineman, and K.~R. Rajagopal, \emph{Mechanical response of polymers---an
  introduction}, Cambridge University Press, Cambridge, 2000.

\bibitem[Karra and
  Rajagopal(2009{\natexlab{a}})]{karra.s.rajagopal.kr:development}
S.~Karra, and K.~R. Rajagopal, \emph{Appl. Mat.} \textbf{54}, 147--176
  (2009{\natexlab{a}}).

\bibitem[Karra and
  Rajagopal(2009{\natexlab{b}})]{karra.s.rajagopal.kr:thermodynamic}
S.~Karra, and K.~R. Rajagopal, \emph{Acta Mech.} \textbf{205}, 105--119
  (2009{\natexlab{b}}).

\bibitem[M\'alek et~al.(2018)]{malek.j.rajagopal.kr.ea:derivation}
J.~M\'alek, K.~R. Rajagopal, and K.~T\r{u}ma, \emph{Fluids} \textbf{3} (2018).

\bibitem[Hron et~al.(2014)]{hron.j.rajagopal.kr.ea:flow}
J.~Hron, K.~R. Rajagopal, and K.~T\r{u}ma, \emph{J. Non-Newton. Fluid Mech.}
  \textbf{210}, 66--77 (2014).

\bibitem[M\'alek
  et~al.(2015{\natexlab{b}})]{malek.j.rajagopal.kr.ea:thermodynamically}
J.~M\'alek, K.~R. Rajagopal, and K.~T\r{u}ma, \emph{Int. J. Pavement Eng.}
  \textbf{16}, 297--314 (2015{\natexlab{b}}).

\bibitem[M\'alek et~al.(2016)]{malek.j.rajagopal.kr.ea:thermodynamically*1}
J.~M\'alek, K.~R. Rajagopal, and K.~T\r{u}ma, \emph{Int. J. Pavement Eng.}
  \textbf{17}, 550--564 (2016).

\bibitem[T\r{u}ma et~al.(2018)]{tuma.k.stein.j.ea:motion}
K.~T\r{u}ma, J.~Stein, V.~Pr\r{u}\v{s}a, and E.~Friedmann, \emph{Appl. Math.
  Comput.} \textbf{335}, 50--64 (2018).

\bibitem[Gurtin et~al.(2010)]{gurtin.me.fried.e.ea:mechanics}
M.~E. Gurtin, E.~Fried, and L.~Anand, \emph{The mechanics and thermodynamics of
  continua}, Cambridge University Press, Cambridge, 2010.

\bibitem[Phan~Thien and Tanner(1977)]{phan-thien.n.tanner.ri:new}
N.~Phan~Thien, and R.~I. Tanner, \emph{J. Non-Newton. Fluid Mech.} \textbf{2},
  353--365 (1977).

\bibitem[Bird et~al.(1980)]{bird.rb.dotson.pj.ea:polymer}
R.~B. Bird, P.~J. Dotson, and N.~L. Johnson, \emph{J. Non-Newton. Fluid Mech.}
  \textbf{7}, 213--235 (1980).

\bibitem[Keunings(1997)]{keunings.r:on}
R.~Keunings, \emph{J. Non-Newton. Fluid Mech.} \textbf{68}, 85--100 (1997).

\bibitem[Phan~Thien(1978)]{phan-thien.n:non-linear}
N.~Phan~Thien, \emph{J. Rheol.} \textbf{22}, 259--283 (1978).

\bibitem[Ngamaramvaranggul and
  Webster(2002)]{ngamaramvaranggul.v.webster.m.f.:simulation}
V.~Ngamaramvaranggul, and M.~F. Webster, \emph{Int. J. Numer. Meth. Fluids}
  \textbf{38}, 677--710 (2002).

\bibitem[Fredrickson(1970)]{fredrickson.ag:model}
A.~G. Fredrickson, \emph{AIChE J.} \textbf{16}, 436--441 (1970).

\bibitem[Bautista et~al.(1999)]{bautista.f.santos.jm.ea:understanding}
F.~Bautista, J.~M. de~Santos, J.~E. Puig, and O.~Manero, \emph{J. Non-Newton.
  Fluid Mech.} \textbf{80}, 93--113 (1999).

\bibitem[Bautista et~al.(2000)]{bautista.f.soltero.jfa.ea:on}
F.~Bautista, J.~F.~A. Soltero, J.~H. P\'erez-L\'opez, J.~E. Puig, and
  O.~Manero, \emph{J. Non-Newton. Fluid Mech.} \textbf{94}, 57--66 (2000).

\bibitem[Bautista et~al.(2002)]{bautista.f.soltero.jfa.ea:irreversible}
F.~Bautista, J.~F.~A. Soltero, E.~R. Mac\'{\i}as, J.~E. Puig, and O.~Manero,
  \emph{J. Phys. Chem. B} \textbf{106}, 13018--13026 (2002).

\bibitem[L\'opez-Aguilar
  et~al.(2016)]{lopez-aguilar.je.webster.mf.ea:comparative}
J.~L\'opez-Aguilar, M.~Webster, H.~Tamaddon-Jahromi, and O.~Manero,
  \emph{Rheol. Acta} \textbf{55}, 197--214 (2016).

\bibitem[Bul\'{\i}\v{c}ek et~al.(2018)]{bulvcek.m.malek.j.ea:pde-analysis}
M.~Bul\'{\i}\v{c}ek, J.~M\'alek, V.~Pr\r{u}\v{s}a, and E.~S\"uli, \enquote{A
  {PDE}-analysis for a class of thermodynamically compatible viscoelastic rate
  type fluids with stress diffusion,} in \emph{Mathematical analysis in fluid
  mechanics: {S}elected recent results}, edited by R.~Danchin, R.~Farwig,
  J.~Neustupa, and P.~Penel, American Mathematical Society, 2018, vol. 710 of
  \emph{Contemporary Mathematics}, pp. 25--53.

\bibitem[Bul{\'{\i}}{\v c}ek et~al.(2018)]{bul-cek.m.feireisl.e.ea:on}
M.~Bul{\'{\i}}{\v c}ek, E.~Feireisl, and J.~M{\'a}lek, \emph{ArXiv e-prints}
  (2018), \eprint{1810.00271}.

\bibitem[Coleman(1970)]{coleman.bd:on}
B.~D. Coleman, \emph{Arch. Ration. Mech. Anal.} \textbf{36}, 1--32 (1970).

\bibitem[Gurtin(1973)]{gurtin.me:thermodynamics*1}
M.~E. Gurtin, \emph{Arch. Ration. Mech. Anal.} \textbf{52}, 93--103 (1973).

\bibitem[Gurtin(1975)]{gurtin.me:thermodynamics}
M.~E. Gurtin, \emph{Arch. Ration. Mech. Anal.} \textbf{59}, 63--96 (1975).

\bibitem[Bul{\'{\i}}{\v c}ek et~al.(2017)]{bul-cek.m.malek.j.ea:thermodynamics}
M.~Bul{\'{\i}}{\v c}ek, J.~M{\'a}lek, and V.~Pr{\r{u}}{\v s}a, \emph{ArXiv
  e-prints}  (2017), \eprint{1709.05968}.

\bibitem[Dostal\'{\i}k et~al.(2018)]{dostalk.m.prusa.v.ea:finite}
M.~Dostal\'{\i}k, V.~{Pr\r{u}\v{s}a}, and K.~{T\r{u}ma}, \emph{ArXiv e-prints}
  (2018), \eprint{1808.03111}.

\end{thebibliography}
\end{document}